\documentclass[12pt]{article}
\usepackage{epsfig,amsmath,amsfonts,amssymb,makeidx,ifthen}
\makeatletter
\@addtoreset{equation}{section}
\makeatother

\makeatletter
\long\def\@makecaption#1#2{{\small
\advance\leftskip1cm
\advance\rightskip1cm
\vskip\abovecaptionskip
\sbox\@tempboxa{#1: #2}%
\ifdim \wd\@tempboxa >\hsize
 #1: #2\par
\else
\global \@minipagefalse
\hb@xt@\hsize{\hfil\box\@tempboxa\hfil}%
\fi
\vskip\belowcaptionskip}}
\makeatother
\def\eq#1\en{\begin{equation}#1\end{equation}}  
\def\eqa#1\ena{\begin{align}#1\end{align}}
\def\eqg#1\eng{\begin{gather}#1\end{gather}}
\newcommand{\lb}[1]{\label{e:#1}}
\newcommand{\rlb}[1]{\eqref{e:#1}} 
\newcommand{\nl}{\notag\\}
\newcommand{\itext}{\notag\intertext}

\newcommand{\sbkt}[1]{\langle#1\rangle}
\newcommand{\bbkt}[1]{\bigl\langle#1\bigr\rangle}
\newcommand{\Bbkt}[1]{\Bigl\langle#1\Bigr\rangle}

\newcommand{\betaon}{\beta_1,\ldots,\beta_n}
\newcommand{\Ga}{\Gamma}
\newcommand{\Gas}{\Gamma^*}
\newcommand{\Gh}{\hat{\Gamma}}
\newcommand{\Ghp}{\hat{\Gamma}_+}
\newcommand{\Ghm}{\hat{\Gamma}_-}
\newcommand{\Gi}{\Gamma_\mathrm{i}}

\newcommand{\ah}{\hat{\alpha}}
\newcommand{\ahs}{\hat{\alpha}_\mathrm{s}}
\newcommand{\ai}{\alpha_\mathrm{i}}
\newcommand{\af}{\alpha_\mathrm{f}}
\newcommand{\ti}{t_\mathrm{i}}
\newcommand{\tf}{t_\mathrm{f}}

\newcommand{\sts}{\mathrm{ss}}
\newcommand{\sG}{\sts\to\Gamma}
\newcommand{\sGs}{\sts\to\Gamma^*}
\newcommand{\Gs}{\Gamma\to\sts}
\newcommand{\Gss}{\Gamma^*\to\sts}
\newcommand{\bsG}[1]{\sbkt{#1}_{\sG}}
\newcommand{\bsGs}[1]{\sbkt{#1}_{\sGs}}
\newcommand{\bGs}[1]{\sbkt{#1}_{\Gs}}
\newcommand{\bGss}[1]{\sbkt{#1}_{\Gss}}
\newcommand{\bpsG}[1]{\sbkt{#1}'_{\sG}}
\newcommand{\bpsGs}[1]{\sbkt{#1}'_{\sGs}}
\newcommand{\bpGs}[1]{\sbkt{#1}'_{\Gs}}
\newcommand{\bpGss}[1]{\sbkt{#1}'_{\Gss}}
\newcommand{\rsa}{\rho^\sts_\alpha}
\newcommand{\rs}{\rho^\sts}
\newcommand{\Thetae}{\Theta^\mathrm{ex}}
\newcommand{\Thetaa}{\Theta_{\ah}}
\newcommand{\Thetaaex}{\Theta_{\ah}^\mathrm{ex}}

\newcommand{\bkta}[1]{\sbkt{#1}^{\ah}}
\newcommand{\Teas}{\sbkt{\Theta^\mathrm{ex}_{\ahs}}^{\ahs}}
\newcommand{\oeo}{O(\epsilon)}
\newcommand{\oes}{O(\epsilon^2)}
\newcommand{\oec}{O(\epsilon^3)}
\newcommand{\Di}{\delta}
\newcommand{\Dei}{\mathit{\Delta}}
\newcommand{\odo}{O(\Di)}
\newcommand{\ods}{O(\Di^2)}
\newcommand{\oeod}{O(\epsilon\Di)}
\newcommand{\oesd}{O(\epsilon^2\Di)}
\newcommand{\oecd}{O(\epsilon^3\Di)}

\newcommand{\calW}{\mathcal{W}}
\newcommand{\Wa}{\calW_{\ah}}
\newcommand{\DGh}{\mathcal{D}\Gh}
\newcommand{\sumn}{\sum_{k=1}^n}
\newcommand{\ca}{(\alpha)}
\newcommand{\cpa}{(\alpha')}
\newcommand{\caeq}{(\alpha_\mathrm{eq})}

\newcommand{\iG}{\int d\Gamma\,}
\newcommand{\rG}{\rho(\Ga)}
\newcommand{\rpG}{\rho'(\Ga)}
\newcommand{\rGs}{\rho(\Gas)}
\newcommand{\rpGs}{\rho'(\Gas)}
\newcommand{\DrG}{\Dei\rho(\Ga)}
\newcommand{\DrGs}{\Dei\rho(\Gas)}
\newcommand{\alphaeq}{\alpha_\mathrm{eq}}
\newcommand{\Ss}{S_\mathrm{sym}}
\newcommand{\DpG}{\Dei\psi(\Gamma)}
\newcommand{\oh}{\frac{1}{2}}
\newcommand{\bn}{\bigskip\noindent}
\newcommand{\deltan}{\eta}

\begin{document}
\noindent
{\bf\large Entropy and Nonlinear Nonequilibrium Thermodynamic Relation for Heat Conducting Steady States}
\par\bigskip

\noindent
Teruhisa S. Komatsu\footnote{
Department of Applied Physics, The University of Tokyo,
 Hongo, Bunkyo, Tokyo 113-8656, Japan
}, 
Naoko Nakagawa\footnote{ 
College of Science, 
 Ibaraki University, Mito, Ibaraki 310-8512, Japan},
Shin-ichi Sasa\footnote{
Department of Pure and Applied Sciences, The University of Tokyo,
 Komaba, Meguro-ku, Tokyo 153-8902, Japan
}, and Hal Tasaki\footnote{
Department of Physics, Gakushuin University, Mejiro, Toshima-ku, Tokyo 171-8588,
 Japan
}

\bigskip

\begin{abstract}
Among various possible routes to extend entropy and thermodynamics to nonequilibrium steady states (NESS), we take the one which is guided by  operational thermodynamics and the Clausius relation.
In our previous study, we derived the extended Clausius relation for NESS, where the heat in the original relation is  replaced by its ``renormalized'' counterpart called the excess heat, and the Gibbs-Shannon expression for the entropy by a new symmetrized Gibbs-Shannon-like expression.
Here we concentrate on Markov processes describing heat conducting systems, and develop a new method for deriving thermodynamic relations.
We first present a new simpler derivation of the extended Clausius relation, and clarify its close relation with the linear response theory.
We then derive a new improved extended Clausius relation with a ``nonlinear nonequilibrium'' contribution  which is written as a correlation between work and heat.
We argue that the ``nonlinear nonequilibrium'' contribution is unavoidable, and is determined uniquely once we accept the (very natural) definition of the excess heat.
Moreover it turns out that to operationally determine the difference in the nonequilibrium entropy to the second order in the temperature difference, one may only use the previous Clausius relation without a nonlinear term or must use the new relation, depending on the operation (i.e., the path in the parameter space).
This peculiar ``twist'' may be a clue to a better understanding of thermodynamics and statistical mechanics of NESS.
\end{abstract}


\tableofcontents

\section{Introduction}
\subsection{Background and motivation}
To develop a universal statistical mechanics for nonequilibrium steady states (NESS) is a major remaining challenge in theoretical physics.
It has been expected since the early days in the research that to pin down a physically natural entropy (or free energy) can be a crucial step in such a project.
It should be realized, however, that the entropy in equilibrium systems plays several essentially different roles; it is a thermodynamic function whose derivatives correspond to physically observable quantities, it is a quantitative measure of which adiabatic process is possible and which is not, it is also a large deviation functional governing the fluctuation of physical quantities.
There is a possibility that this ``degeneracy'' is an accident observed only in equilibrium states, and the degeneracy is immediately lifted when one goes out of equilibrium.
If this is the case we shall encounter more than one ``nonequilibrium entropies'' each of which characterizing different physical aspect of a NESS.
It will then be important to clarify what physics of NESS is represented by which extension of entropy.
Among various promising attempts, we refer to the phenomenological consideration in \cite{OP} which tried to extend the notion of adiabatic accessibility, and recent explicit characterization of the large deviation functional in \cite{DLS,Rome,BD}.

In our own attempt to extend the notion of entropy to NESS \cite{1}, we have concentrated on the aspect which gave  birth to the concept of entropy, namely, the Clausius relation in thermodynamics.
We started from a microscopic description of a heat conducting NESS, and showed that a very natural generalization of the Clausius relation, in which the heat is replaced with the ``excess heat'', is indeed possible when the ``order of nonequilibrium'' $\epsilon$ is sufficiently small.
This was a realization of the early phenomenological discussions in \cite{La,OP}, and extension of the similar result in \cite{Ruelle} for models with Gaussian thermostat.
We also found that the microscopic representation of the entropy differs from the traditional  Gibbs-Shannon form, and requires further symmetrization with respect to time-reversal transformation (see \rlb{Sa} below).

In the present paper, where we treat a class of Markov processes describing heat conducting systems, we shall go one step further to show a new thermodynamic relation that has the same entropy as in \cite{1}, but contains an extra correction term which is intrinsically ``nonlinear nonequilibrium'' and has no counterpart in the original Clausius relation.
Although one might worry that showing new relations with new nonlinear terms is a rather arbitrary uncontrolled attempt, we argue that {\em this extension is mandatory and unique}\/ if we regard the extended Clausius relation of \cite{1} as a starting point.
Moreover we shall see that to determine the entropy to the second order in the ``order of nonequilibrium'' $\epsilon$, one may only use the naive extended Clausius relation \eqref{e:Cl} or must use the new relation \eqref{e:new} with an extra term, depending on the operation under consideration.
This fact suggests that {\em there is a delicate but unavoidable ``twist'' in operational thermodynamics for NESS}.
We believe that one must understand the true nature of this twist before developing a satisfactory statistical theory of NESS.

\subsection{Brief summary of the setting and the results}
\label{s:excess}
Before discussing the details, let us present a brief and informal summary of our approach to NESS and major results in the present paper.
We shall in particular explain the important notions of excess heat and excess entropy production (in the baths).

\begin{figure}[btp]
\begin{center}
\includegraphics[width=14cm]{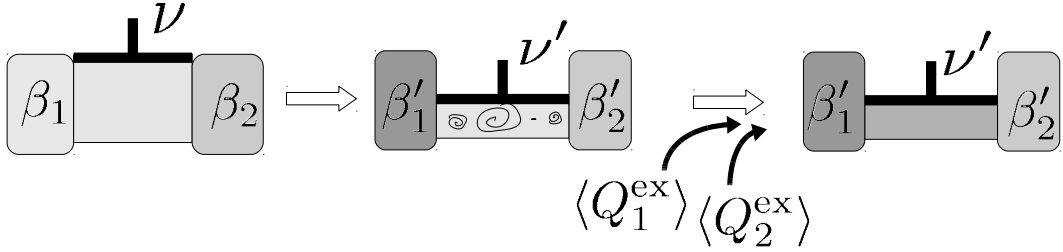}
\end{center}
\caption[dummy]{
Thermodynamic operation in a NESS.
Left: One starts from the NESS  with $\beta_1$, $\beta_2$, and $\nu$.
Middle: At a certain moment an external agent abruptly changes the inverse temperatures of the baths and the parameter.  The system is no longer in a steady state.
Right:  After a sufficiently long time, the system relaxes to a new NESS.
In the extended Clausius relation \rlb{exCl0}, the excess heat or the ``renormalized'' heat ($\bbkt{Q_1^\mathrm{ex}},\bbkt{Q_2^\mathrm{ex}}$)
plays an essential role since the ``bare'' heat diverges in time.
(see also Fig.\ref{fig:Qex}.)
}
\label{fig:OPNESS}
\end{figure}

Let us focus on the simplest nontrivial example, i.e., a system (of, say, a fluid) attached to two heat baths with inverse temperatures $\beta_1$ and $\beta_2$.
We assume that the system is characterized by a parameter (or a set of parameters) $\nu$ that can be controlled by an outside agent.
A typical example of $\nu$ is the volume of the system.
See the left-most figure in Fig.~\ref{fig:OPNESS}.

\paragraph*{Stepwise operation in equilibrium:}
We start with the case where  the inverse temperatures of the two bath are both equal to $\beta$.
Suppose that we fix both $\beta$ and $\nu$.
Then after a sufficiently long time,  the system reaches a unique equilibrium state characterized by $\beta$ and $\nu$.

\begin{figure}[btp]
\begin{center}
\includegraphics[width=14cm]{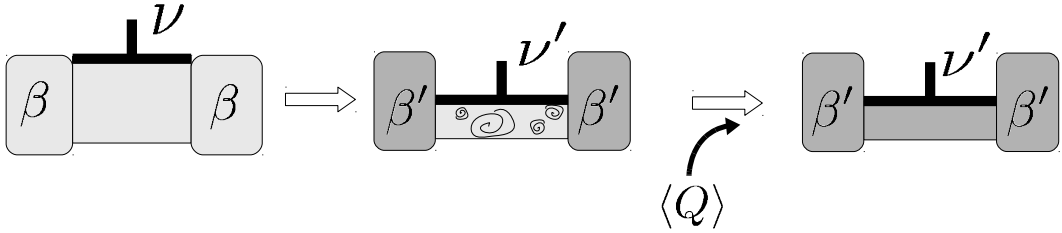}
\end{center}
\caption[dummy]{
Thermodynamic operation in an equilibrium state.
Left: One starts from the equilibrium state with the inverse temperature $\beta$ and the parameter $\nu$.
Middle: At a certain moment an external agent abruptly changes the inverse temperature of the baths and the parameter.  The system is no longer in equilibrium.
Right:  After a sufficiently long time, the system relaxes to the new equilibrium corresponding to $\beta'$ and $\nu'$.
The heat $\bbkt{Q}$ absorbed by the system during the relaxation plays an essential role in the Clausius relation \rlb{Cl0}.
}
\label{fig:OPEQ}
\end{figure}

We consider a thermodynamic operation to the equilibrium state.
Suppose that an external agent instantaneously changes the inverse temperature (of both the baths) from $\beta$ to $\beta'$ and the parameter from $\nu$ to $\nu'$.
The system is no longer in equilibrium after this sudden operation, but finally relaxes into the new equilibrium state corresponding to $\beta'$ and $\nu'$ after a sufficiently long time.
Let us denote by $Q$ the total heat that flows into the system from the heat baths during the relaxation process,
and by $\bbkt{Q}$ its average over possible paths (i.e., histories).
See Fig.~\ref{fig:OPEQ}.
Then the celebrated Clausius relation states that
\eq
S_\mathrm{eq}(\beta';\nu')-S_\mathrm{eq}(\beta;\nu)=\beta\, \bbkt{Q}+\ods,
\lb{Cl0}
\en
where $S_\mathrm{eq}(\beta;\nu)$ is the entropy of the equilibrium state,
and $\Di$ is the dimensionless measure of the changes $\beta'-\beta$ and $\nu'-\nu$.
The error term $\ods$ simply reflects the stepwise nature of the operation, and is not essential.

\paragraph*{Stepwise operation in NESS:}
Our goal is to extend the previous consideration for equilibrium states to NESS.
Let $\beta_1\ne\beta_2$, and fix $\beta_1$, $\beta_2$ and $\nu$.
Then again after a sufficiently long time, the system is expected to settle to a unique NESS\footnote{
We are assuming that the temperature difference $|\beta_1-\beta_2|$ is sufficiently small.
}, which exhibits no macroscopic changes but has a nonvanishing heat current.

We consider a thermodynamic operation similar to the equilibrium case.
Suppose that the system is in the NESS characterized by $\beta_1$, $\beta_2$, and $\nu$.
Then the external agent instantaneously changes the inverse temperatures of the baths from $\beta_1$ to $\beta'_1$ and $\beta_2$ to $\beta'_2$, respectively, and the parameter from $\nu$ to $\nu'$.
The system is no longer in a steady state after the change, but will eventually relax to a new NESS characterized by $\beta'_1$, $\beta'_2$, and $\nu'$.
We want to examine if there is a relation like the Clausius relation \rlb{Cl0} which holds in this situation.

Let us now see that a naive extension of \rlb{Cl0} is just impossible.
Recall that, in the equilibrium case, the quantity $-\beta\, \bbkt{Q}$ is precisely the increase of entropy in the baths during the relaxation process.
Thus the Clausius relation  \rlb{Cl0} simply states that the change in the total entropy of the system and the baths is at most of $\ods$.
In a NESS, on the contrary, there is a constant increase in the entropy of the baths even when one makes no operation.
To see this consider a NESS with $\beta_1$, $\beta_2$, and $\nu$, and suppose that $\beta_1<\beta_2$.
Then there is a constant heat current $J_1^\mathrm{ss}>0$ from the first heat bath to the system and the same amount from the system to the second bath.
One then easily finds that the total entropy in the baths grows linearly in time with a constant rate $(\beta_2-\beta_1)\,J_1^\mathrm{ss}>0$.
Thus, in the operation to NESS, a naive counterpart of the quantity $-\beta\, \bbkt{Q}$ is growing constantly before the operation and keeps on growing constantly after the operation.
There is no hope that a thermodynamic relation like \rlb{Cl0} holds.

\begin{figure}[btp]
\begin{center}
\includegraphics[width=10cm]{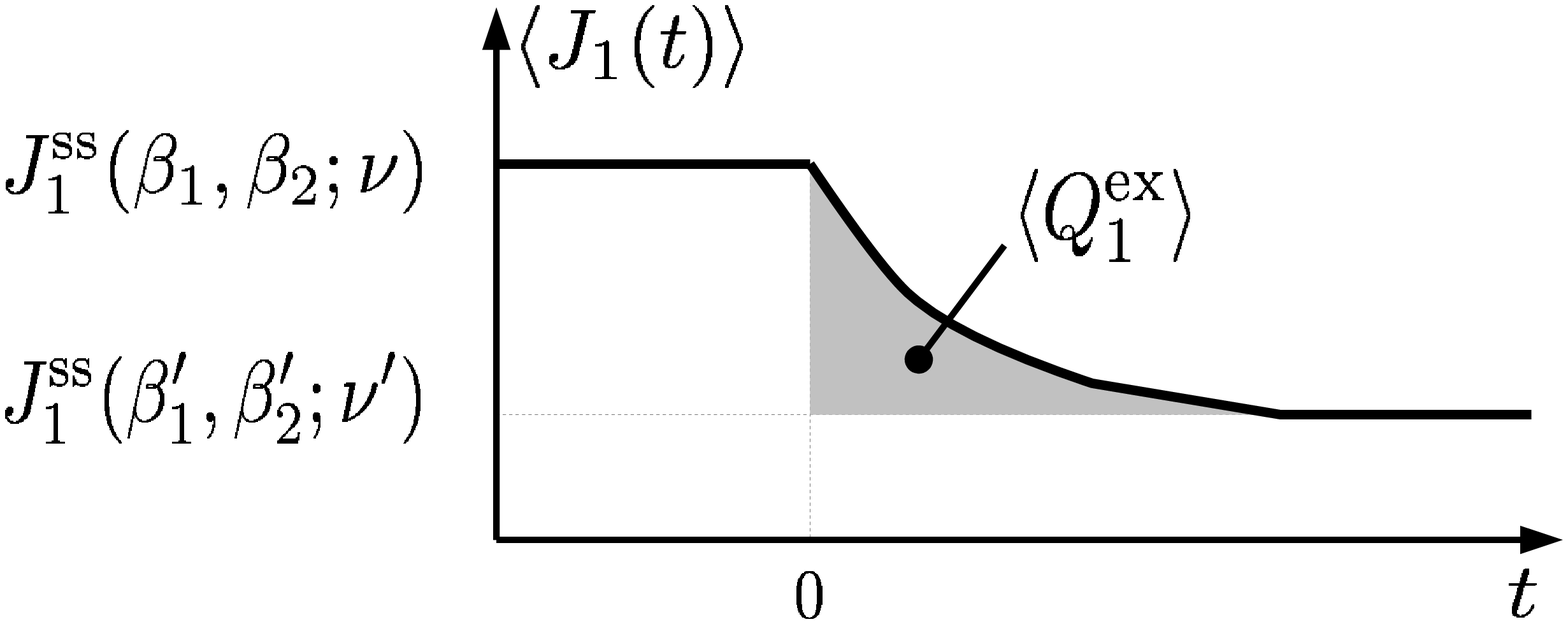}
\end{center}
\caption[dummy]{
$\sbkt{J_1(t)}$ denotes the averaged heat current from the first bath to the system at time $t$.
The area of the region below the solid line corresponds to the total heat from the bath to the system, which diverges in time.
If one subtracts the heat necessary to maintain the NESS, one gets the shaded area which corresponds to the excess heat $\bbkt{Q_1^\mathrm{ex}}$.
This is a very natural strategy to ``renormalize'' diverging ``bare'' heat to get a finite quantity.
}
\label{fig:Qex}
\end{figure}

\paragraph*{Excess heat, excess entropy production:}
To overcome this difficulty of divergence, we shall consider a properly ``renormalized'' version of the diverging ``bare'' quantity.
In the present situation a natural
procedure of the renormalization is to subtract ``house-keeping heat'' from the diverging ``bare heat'' in the spirit of Oono and Paniconi \cite{OP}.

To be precise let $J_k(t)$ ($k=1,2$) be the heat current that flows from the $k$-th bath to the system at time $t$, and $\bbkt{J_k(t)}$ be its average.
The total heat $Q_k$ from the $k$-th bath is given by the integral $\int dt\,{J_k(t)}$, whose average $\bbkt{Q_k}$ clearly diverges as the range of the integral is enlarged.

Let $J_k^\mathrm{ss}(\beta_1,\beta_2;\nu)$ be the heat current from the $k$-th bath to the system\footnote{
We of course have $J_1^\mathrm{ss}(\beta_1,\beta_2;\nu)=-J_2^\mathrm{ss}(\beta_1,\beta_2;\nu)$ from the energy conservation.
} expected in the NESS characterized by $\beta_1$, $\beta_2$, and $\nu$.
Suppose that we make the sudden change of the parameters at time $t=0$.
Then the current needed to maintain the NESS is $J_k^\mathrm{ss}(\beta_1,\beta_2;\nu)$ for $t\le0$, and  $J_k^\mathrm{ss}(\beta'_1,\beta'_2;\nu')$ for $t\ge0$.
We define the excess heat (or the renormalized heat) by substituting these steady heat currents (which correspond to the ``house keeping heat'' in \cite{OP}) from the actual current $\bbkt{J_k(t)}$.
Thus the excess current is given by $\bbkt{J_k(t)}-J_k^\mathrm{ss}(\beta_1,\beta_2;\nu)$ for $t\le0$, and  $\bbkt{J_k(t)}-J_k^\mathrm{ss}(\beta'_1,\beta'_2;\nu')$ for $t\ge0$.
The excess (or renormalized) heat $ Q_k^\mathrm{ex}$, which is expected to characterize intrinsic heat transfer caused by the operation, is obtained by integrating this excess current.
By noting that $\bbkt{J_k(t)}=J_k^\mathrm{ss}(\beta_1,\beta_2;\nu)$ for $t\le0$, we have
\eq
\bbkt{Q_k^\mathrm{ex}}:=\int_0^\infty dt\,\Bigl\{\bbkt{J_k(t)}-J_k^\mathrm{ss}(\beta'_1,\beta'_2;\nu')\Bigr\},
\en
which, unlike the total heat, is expected to be finite.
See Fig.~\ref{fig:Qex}.

\paragraph*{Extended Clausius relation and its improvement:}
Our finding in \cite{1} is that there is a sensible extension of the Clausius relation \rlb{Cl0} in which the heat is replaced by the excess heat.
More precisely we showed the extended Clausius relation 
\eq
S(\beta'_1,\beta'_2;\nu')-S(\beta_1,\beta_2;\nu)=\beta_1\, \bbkt{Q_1^\mathrm{ex}}+\beta_2\, \bbkt{Q_2^\mathrm{ex}}+\oesd+\ods,
\lb{exCl0}
\en
where the nonequilibrium entropy $S(\beta_1,\beta_2;\nu)$ is a well defined function of $\beta_1$, $\beta_2$, and $\nu$, and coincides with the equilibrium entropy if $\beta_1=\beta_2$.
The quantity $-(\beta_1\, \bbkt{Q_1^\mathrm{ex}}+\beta_2\, \bbkt{Q_2^\mathrm{ex}})$ in the right-hand side of \rlb{exCl0}, which we call the {\em excess entropy production}\/ in the baths, plays an essential role in our thermodynamic approach to NESS.
Moreover we found that the nonequilibrium entropy $S(\beta_1,\beta_2;\nu)$ can be written as a symmetrized Shannon-Gibbs entropy \rlb{Sa} in terms of the probability distribution of the NESS.

As we have already announced in the previous section, we derive in the present paper an improved ``nonlinear'' version of the extended Clausius relation \rlb{exCl0}.
In the simplest case where only the parameter is changed abruptly from $\nu$ to $\nu'$, and the inverse temperatures do not change, the new relation reads
\eqa
S(\beta'_1,\beta'_2;\nu')-S(\beta_1,\beta_2;\nu)=&\beta_1\, \bbkt{Q_1^\mathrm{ex}}+\beta_2\, \bbkt{Q_2^\mathrm{ex}}-\frac{\beta_1+\beta_2}{4}\bbkt{W;(\beta_1Q_1+\beta_2Q_2)}
\nl&+\oecd+\ods.
\ena
Here $W$ is the total work done by the agent during the operation, and $\bbkt{A;B}=\bbkt{AB}-\bbkt{A}\,\bbkt{B}$ stands for the truncated correlation.

Implications of these results on thermodynamics of NESS are discussed carefully in section~\ref{s:TDtwist}.

\section{Main results}
\subsection{Setup and basic definitions}
\paragraph*{System and basic parameters:}
For simplicity, we treat the same class of models as in \cite{1}, namely, a heat conducting system with controllable parameters (such as the volume).
More precisely we consider a classical system with  $N$ particles  
whose coordinates are collectively denoted as 
\eq
\Ga=(\mathbf{p}_1,\ldots,\mathbf{p}_N;\mathbf{r}_1,\ldots,\mathbf{r}_N),
\en
where $\mathbf{p}_j$ and $\mathbf{r}_j$ are the momentum and the position, respectively, of the $j$-th particle.
The time-reversal of a state $\Ga$ is 
\eq
\Gas
=(-\mathbf{p}_1,\ldots,-\mathbf{p}_N;\mathbf{r}_1,\ldots,\mathbf{r}_N).
\en
The Hamiltonian $H_\nu(\Gamma)$ has a controllable parameter $\nu$, and satisfies the time reversal symmetry $H_\nu(\Gamma)=H_\nu(\Gamma^*)$.
The system is attached to $n$ distinct heat baths with (generally different) inverse temperatures $\betaon$.
We denote the parameters of the model collectively as 
\eq
\alpha=(\betaon;\nu).
\lb{alpha}
\en

The time evolution of the system is basically governed by the deterministic Newtonian dynamics corresponding to the Hamiltonian $H_\nu(\Ga)$, but the effects of the heat baths are taken into account by a suitable Markovian dynamics\footnote{
The standard choice is the  Langevin dynamics or the thermal wall \cite{LF}.
It is essential that the dynamics is Markovian and satisfies the local detailed balance condition \cite{BL}.
}.
Our results do not depend on details how one implements the heat baths. 

We assume that the system reaches a unique NESS after developing for a sufficiently long time under a given set of parameters $\alpha$.
By $\rsa(\Gamma)$ we denote the probability distribution in the NESS.

We introduce the dimensionless measure of the ``order of nonequilibrium'' corresponding to $\alpha=(\betaon;\nu)$ as
\eq
\epsilon:=\max_{k\in\{1,\ldots,n\}}\frac{|\beta_k-\beta|}{\beta},
\lb{ep}
\en
where the ``reference inverse temperature'' $\beta$ is a quantity close to $\betaon$, which can be chosen rather arbitrarily.
For concreteness we here set
\eq
\beta:=\frac{1}{n}\sum_{k=1}^n\beta_k.
\lb{bref}
\en

\paragraph*{Operation and path average:}
It is crucial for us to consider an operation by an external agent.
During the time interval $[\ti,\tf]$, the agent controls the model parameters according to a pre-fixed protocol represented as a function 
\eq
\ah=(\alpha(t))_{t\in[\ti,\tf]}.
\lb{patha}
\en
We write the initial and the final values as $\ai=\alpha(\ti)$ and $\af=\alpha(\tf)$.
We also denote by $\Gh=(\Gamma(t))_{t\in[\ti,\tf]}$ a possible path of the system in the same time interval $[\ti,\tf]$.
We denote by $\Wa(\Gh)$ the stochastic ``weight'' for a path $\Gh$, which is properly normalized so that
\eq
\int\DGh\,\delta(\Gamma(\ti)-\Gi)\,\Wa(\Gh)=1
\en 
holds for any initial state $\Gi$, where the ``integral'' is over all the possible paths\footnote{
It is known that such a ``path integral'' is ill-defined (or requires an extra care in definition) for certain continuous-time stochastic processes.
In such a case we first consider a discrete-time approximation of the process, where the corresponding path integral is well-defined.
After deriving the desired relations, we can take the continuum limit to recover the original process.
In what follows, we shall not make this procedure explicit.
} $\Gh$.

For an arbitrary function $f(\Gh)$ of a path, we define its average in the protocol $\ah$ as\footnote{
When we define $A$ in terms of $B$, we write $A:=B$ or $B=:A$.
} 
\eq
\bkta{f}:=\int\DGh\,\rs_{\ai}(\Gamma(\ti))\,\Wa(\Gh)\,f(\Gh).
\lb{Fa}
\en
Note that we have summed over all the paths, assuming that the system in the NESS with $\ai$ at time $t=\ti$.

\paragraph*{Entropy production and excess entropy production:}
Let $J_k(\Gh;t)$ be the heat flux that flows into the system from the $k$-th bath at time $t$ in a path $\Gh$.
The most important quantity in the present theory is the {\em total entropy production}\/ in the baths defined as
\eq
\Thetaa(\Gh):=-\int_{\ti}^{\tf}dt\sumn\beta_k(t)\,J_k(\Gh;t).
\lb{Theta}
\en
We also introduce the {\em excess entropy production}\/ in the baths
\eq
\Thetaaex(\Gh):=
-\int_{\ti}^{\tf}dt\sumn\beta_k(t)\bigl\{J_k(\Gh;t)-J^\sts_k(\alpha(t))\bigr\},
\lb{Thetae}
\en
where $J^\sts_k(\alpha)$ denotes the steady heat current in NESS with fixed $\alpha$.
Note that, for the stepwise operation considered in Section~\ref{s:excess}, $\Thetaaex(\Gh)$ coincides with the quantity $-(\beta_1\, Q_1^\mathrm{ex}+\beta_2\, Q_2^\mathrm{ex})$ in \rlb{exCl0}.
One can also say that the excess entropy production \rlb{Thetae} is a ``renormalized''version of the ``bare'' entropy production \rlb{Theta} which grows almost linearly in $\tf-\ti$.

\subsection{Clausius relation and its extensions}
\label{s:TD}
\paragraph*{Original Clausius relation:}
Before dealing with NESS, we consider the corresponding equilibrium problem where all the baths have the same temperature.
Let $\ah$ describe an arbitrary quasi-static {\em equilibrium}\/ operation, i.e.,  $\alpha(t)$ is an arbitrary slowly varying function with $\beta_1(t)=\cdots=\beta_n(t)$.
Then the classical Clausius relation in its most general form reads
\eq
S_\mathrm{eq}(\af)-S_\mathrm{eq}(\ai)=-\bkta{\Thetaa},
\lb{oldCl}
\en
which is the integrated version of \rlb{Cl0}.
The equilibrium entropy $S_\mathrm{eq}(\alpha)$ has a well-known microscopic expression
\eq
S_\mathrm{eq}(\alpha)=-\iG\rho^\mathrm{eq}_\alpha(\Ga)\,\log\rho^\mathrm{eq}_\alpha(\Ga).
\lb{Seq}
\en
Here 
\eq
\rho^\mathrm{eq}_\alpha(\Ga)=\frac{e^{-\beta\,H_\nu(\Ga)}}{Z_\nu(\beta)}
\lb{req}
\en
is the probability distribution in the equilibrium state with $\alpha$, where the partition function $Z_\nu(\beta)$ is determined by the normalization condition $\iG\rho^\mathrm{eq}_\alpha(\Ga)=1$.

Our goal is to find natural extensions to NESS of the Clausius relation \rlb{oldCl} and the corresponding expression  of the entropy like the Gibbs-Shannon expression \rlb{Seq}.

\paragraph*{Extended Clausius relation for NESS:}
As we noted in section~\ref{s:excess}, a key for the extension is to replace a diverging quantity with a ``renormalized'' version.
In  \cite{1} we have shown for an arbitrary quasi-static protocol $\ah$ that
\eq
\Ss(\af)-\Ss(\ai)=-\bkta{\Thetaaex}+\oesd,
\lb{Cl}
\en
where the entropy production \rlb{Theta} in the original relation \rlb{oldCl} is replaced with the excess entropy production \rlb{Thetae}.
Here $\epsilon$ is the ``order of nonequilibrium'' defined as in \rlb{ep}, and $\Di$ is the dimensionless measure of the change in the parameters\footnote{
\label{fn:AC}
If $\alpha(t)$ varies monotonically, one  can simply take the difference $\af-\ai$ (and then properly normalize each component to make the result dimensionless).
For a more general function $\alpha(t)$, one has to consider the ``accumulated change'' such as $\Dei\nu:=\int_{\ti}^{\tf}dt\,|\dot{\nu}(t)|$.
} $\ah$.
The nonequilibrium entropy $\Ss(\alpha)$ is written in terms of  the probability distribution $\rsa(\Ga)$ of the NESS as
\eq
\Ss(\alpha)=-\frac{1}{2}\int d\Ga\,\rsa(\Ga)\bigl\{\log\rsa(\Ga)+\log\rsa(\Gas)\bigr\}.
\lb{Sa}
\en
The expression \rlb{Sa} is similar to the Gibbs-Shannon entropy \rlb{Seq}, but has an extra symmetrization with respect to the time reversal.
Note that for any probability distribution which is invariant under the time-reversal, our symmetrized entropy \rlb{Sa} coincides with the Gibbs-Shannon entropy \rlb{Seq}.
In particular we have $S_\mathrm{eq}(\alpha)=\Ss(\alpha)$ for an equilibrium $\alpha$.

The relation \rlb{Cl} was first derived by Ruelle  \cite{Ruelle} for a class of systems with thermostat, but the expression \rlb{Sa} first appeared in \cite{1}.

Let us investigate the extended Clausius relation \rlb{Cl} from the view point of operational thermodynamics, where one examines thermodynamic quantities through various processes which are experimentally realizable (in principle).
We then find that, by using the relation \rlb{Cl}, one can determine the nonequilibrium entropy $\Ss(\alpha)$ with the precision of $\oes$ from a thermodynamic measurement as follows.
For a given nonequilibrium $\alpha=(\betaon;\nu)$, one defines its equilibrium counterpart by $\alphaeq:=(\beta,\ldots,\beta;\nu)$ where $\beta$ is the reference inverse temperature defined as \rlb{bref}.
One then takes $\ah$ as a quasi-static protocol which follows the straight path\footnote{
More precisely, we set $\beta_j(t)=\beta+(\beta_j-\beta)(t-\ti)/(\tf-\ti)$ and $\nu(t)=\nu$ for $t\in[\ti,\tf]$.
} from  $\alphaeq$ to $\alpha$.
Then the extended Clausius relation \rlb{Cl} implies
\eq
\Ss(\alpha)=S_\mathrm{eq}(\alphaeq)-\bkta{\Thetaaex}+\oec,
\lb{SSeq}
\en
because one has $\Di=\oeo$ in this protocol.
This expression was used in \cite{KNSTI} for a numerical evaluation of the entropy in a NESS.

The fact that we can determine $\Ss(\alpha)$ to the precision of $\oes$ might suggest that we are entering the regime of ``nonlinear nonequilibrium.''
The truth, however, is that the extended Clausius relation \rlb{Cl} itself can  be fully understood in terms of the physics of linear response.
In fact the derivation of \rlb{Cl} in \cite{1} was rather involved, and we did not realize its direct relation to the linear response theory.
In the present paper, we shall present a much clearer derivation of \rlb{Cl}, which is essentially based on the linear response form \rlb{LR} of the probability distribution in NESS.

\bigskip\noindent
{\em Remark:}\/
For the above mentioned reason, we have made in \cite{1} a premature conclusion that the extended Clausius relation \rlb{Cl} goes beyond the linear response theory.
We also remark here that the extended Gibbs relation mentioned in \cite{1} should be reexamined carefully.

\paragraph*{New ``nonlinear nonequilibrium" Clausius relation for NESS:}

The extended Clausius relation \rlb{Cl} has an error term $\oesd$ which is absent in the original Clausius relation \rlb{oldCl}.
Although one might hope that the error may be reduced by a better derivation, we know from examples that there indeed exists a nonvanishing error of this order.
This is an unfortunate fact since \rlb{Cl} seems to be the most natural nonequilibrium extension of the Clausius relation \rlb{oldCl}.
All that we can hope is to derive  improved relations with smaller error terms\footnote{
One may also think about a better method of ``renormalizing'' the divergent entropy production.
It seems that the present prescription is the most natural one, but the naturalness may not guarantee that it is the ``right'' choice.
}.

The main results of the present paper are ``nonlinear nonequilibrium'' improvements \rlb{new}, \rlb{newnu} of the extended Clausius relation \rlb{Cl}.
The improved relations  contain an extra nonlinear term which involves the correlation between  heat and  energy.
While the extended Clausius relation \rlb{Cl} is based on the linear response formula \rlb{LR}, the improved relations \rlb{new}, \rlb{newnu} are derived from the nonlinear representation \rlb{KN} of the probability distribution of NESS which was derived by  two of us (T.S.K. and N.N.) in \cite{KN}.
See also \cite{JSP}.

Now let us describe the new relation.
To characterize the amount of change in the protocol $\ah$, we introduce two dimensionless quantities $\Di_\mathrm{o}$ and $\Di_\mathrm{r}$.
Here  $\Di_\mathrm{o}$ denotes the sum of the accumulated changes (see footnote~\ref{fn:AC}) in the parameters
$\nu(t)$ and $\beta(t):=n^{-1}\sum_{k=1}^{n}\beta_k(t)$, both suitably normalized to be dimensionless.
On the other hand $\Di_\mathrm{r}$ denotes the amount of change in the degree of nonequilibrium.
More precisely it is the sum of the  accumulated changes
 in the relative inverse temperatures $|\beta_k(t)-\beta(t)|$ for $k=1,\ldots,n$.
 
For an arbitrary quasi-static protocol $\ah$, we shall derive
\eq
\Ss(\af)-\Ss(\ai)
=-\bkta{\Thetaaex}+
\frac{1}{2}\bkta{\tilde{W}_{\ah};\Thetaa}+O(\epsilon^2\Di_\mathrm{r})+O(\epsilon^3\Di_\mathrm{o}),
\lb{new}
\en
which is an improvement\footnote{
We can also show that $\bkta{\tilde{W}_{\ah};\Thetaa}=O(\epsilon^2\Di_\mathrm{o})$.
}
 of \rlb{Cl} since one normally has\footnote{
One can indeed make $\Di_\mathrm{r}$ large by, for example, letting $\beta_k(t)$ oscillate rapidly in time.
We here assume that everything is changed smoothly.
} $\Di_\mathrm{r}\le\oeo$.
Here $\sbkt{A;B}=\sbkt{AB}-\sbkt{A}\sbkt{B}$ denotes the truncated correlation, and 
\eq
\tilde{W}_{\ah}:=\int_{\ti}^{\tf}dt\,\frac{d}{ds}\bigl\{\beta(s)\,H_{\nu(s)}(t)\bigr\}\Bigr|_{s=t}
\en
is a quantity which is related to (but different from) the work.  See \rlb{Work}.
Here $\beta(t)$ is the reference inverse temperature \rlb{bref} at time $t$, and the ``order of nonequilibrium'' is defined as
\eq
\epsilon:=\mathop{\max_{k\in\{1,\ldots,n\}}}_{t\in[\ti,\tf]}\frac{|\beta_k(t)-\beta(t)|}{\beta(t)}
\lb{ep2}
\en
according to \rlb{ep}.

We remark here that the correlation term $\bkta{\tilde{W}_{\ah};\Thetaa}$ in \rlb{new} is an intrinsically nonequilibrium contribution.
Although both $\bkta{\tilde{W}_{\ah}}$ and $\bkta{\Thetaa}$ are nonvanishing in a general equilibrium process, there truncated correlation is exactly vanishing.
See \rlb{corvan}. 
We expect on the other hand that the correlation is nonvanishing whenever there is a nonequilibrium correction to the probability distribution of NESS.

For a restricted class of quasi-static protocols where $\betaon$ are fixed constants and only $\nu$ varies, the relation \rlb{new} becomes
\eq
\Ss(\af)-\Ss(\ai)
=-\bkta{\Thetaaex}+
\frac{\beta}{2}\bkta{W_{\ah};\Thetaa}+O(\epsilon^3\Di_\mathrm{o}).
\lb{newnu}
\en
Here the quantity
\eq
W_{\ah}(\Gh):=
\int_{\ti}^{\tf}dt\,\frac{d}{ds}\bigl\{H_{\nu(s)}(t)\bigr\}\Bigr|_{s=t}
=\int_{\ti}^{\tf}dt~\dot\nu(t)\,\frac{\partial H_{\nu}(\Gamma(t))}{\partial\nu}\Bigr|_{\nu=\nu(t)}
\lb{Work}
\en
is nothing but the work done by the external agent who operates the parameter $\nu$.

\begin{figure}[btp]
\begin{center}
\includegraphics[width=7.cm]{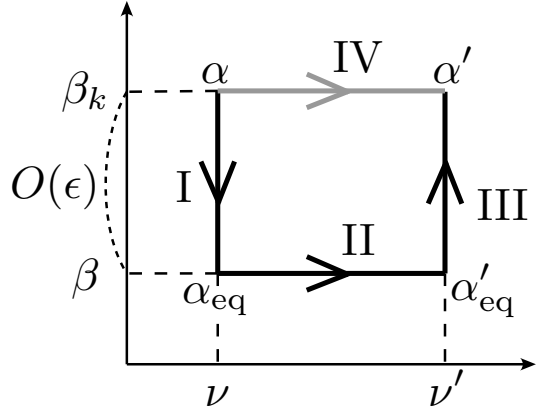}
\end{center}
\caption[dummy]{
The paths in the parameter space discussed in the text, where
$\beta_k$ is the representative of $\betaon$.
Suppose one wants to determine the entropy difference $\Ss(\alpha')-\Ss(\alpha)$ with the precision of $\oes$, by using thermodynamic measurements.
One possibility is to go thorough the indirect paths I, II, II, and use the extended Clausius relation \rlb{Cl} as well as the traditional Clausius relation \rlb{oldCl}.
When one goes through the direct path IV, however, one has to invoke the new ``nonlinear nonequilibrium'' extended Clausius relation \rlb{newnu}.
}
\label{fig:paths}
\end{figure}

\subsection{Unavoidable ``twist'' in thermodynamics of NESS}
\label{s:TDtwist}
It may not be desirable, at least from a practical point of view, that a thermodynamic relation contains a correlation like $\bkta{W_{\ah};\Thetaa}$.
But we shall now argue that this term is unavoidable, and clearly indicates the existence of a delicate ``twist'' in thermodynamics for NESS.

Let us consider the following example.
Take two parameter values $\nu$, $\nu'$ where $\nu-\nu'$ may not be small, and let
\eqa
&\alpha=(\betaon;\nu), \quad \alpha'=(\betaon;\nu')\nl
&\alphaeq=(\beta,\ldots,\beta;\nu), \quad\alphaeq'=(\beta,\ldots,\beta;\nu')
\ena
with $\beta$ as in \rlb{bref}.
See Fig.~\ref{fig:paths}.
Suppose that one wishes to determine the entropy difference $\Ss(\alpha')-\Ss(\alpha)$ to the precision of $\oes$ by using thermodynamic measurements.
An indirect but clever way is to write 
\eqa
\Ss(\alpha')-\Ss(\alpha)=&\{\Ss(\alpha')-\Ss(\alphaeq')\}\nl&+\{\Ss(\alphaeq')-\Ss(\alphaeq)\}
+\{\Ss(\alphaeq)-\Ss(\alpha)\}
\ena
and determine the three differences separately (paths III , II and I in Fig.~\ref{fig:paths}).
To determine $\Ss(\alphaeq)-\Ss(\alpha)$ and $\Ss(\alpha')-\Ss(\alphaeq')$, it suffices to use the (older) extended Clausius relation \rlb{Cl}, whose error term is $\oec$ since one here has $\Di=\oeo$.
To determine $\Ss(\alphaeq')-\Ss(\alphaeq)=S_\mathrm{eq}(\alphaeq')-S_\mathrm{eq}(\alphaeq)$, one simply uses the standard Clausius relation \rlb{oldCl}, which is error free.
Note that all the three determinations involve only direct measurements of heat currents and make use of the standard Clausius relation \rlb{oldCl} or its natural extension\footnote{
One can of course use the extended relations \rlb{new}, which is stronger, to get basically the same result.
} \rlb{Cl}.
It may be reasonable to conclude that the entropy difference $\Ss(\alpha')-\Ss(\alpha)$ (at least to this precision) is a thermodynamically natural quantity to look at.

One may also determine $\Ss(\alpha')-\Ss(\alpha)$ by using the direct path from $\alpha$ to $\alpha'$  in which $\betaon$ are all fixed  and $\nu$ is varied from $\nu$ to $\nu'$
(path IV in Fig.~\ref{fig:paths}).
Note that the (older) extended Clausius relation \rlb{Cl} contains the error of $\oes$ since $\Di$ is not small.
To achieve the same precision of $\oes$ in this case, one {\em must}\/ use the new relation \rlb{new} (or \rlb{newnu} in this particular case) which contains the correlation between work and heat.
It should be stressed  that the left-hand side of \rlb{newnu} is known (from the above consideration) to be a natural quantity to look at, but one must include the ``nonlinear'' correlation term in order to recover the same precision attained in the measurements along the three indirect paths I, II, and III.

Obviously similar consideration applies to much more general situations, where one uses the most general ``nonlinear nonequilibrium'' Clausius relation \rlb{new}.
For example think about determining, again with the precision of $\oes$, the entropy difference between two NESS with $\alpha=(\beta_1,\beta_2,\ldots,\beta_n;\nu)$ and $\alpha''=(\beta_1+\Dei\beta,\beta_2+\Dei\beta,\ldots,\beta_n+\Dei\beta;\nu)$ where $\Dei\beta$ may not be small.
When one considers three paths which go through equilibrium states it suffices to use the extended Clausius relation \rlb{Cl} and the standard Clausius relation \rlb{oldCl}.
If one considers the direct path, on the other hand, one must use the new Clausius relation \rlb{new}.

The above observations, first of all, show that the correlation term in \rlb{new} is  mandatory, and \rlb{new} is essentially the unique reasonable ``nonlinear nonequilibrium'' extension of the Clausius relation \rlb{oldCl}.
It also suggests that there is a delicate ``twist'' in thermodynamics of NESS.
More precisely, we have found that {\em one may use only the simple extended Clausius relation \rlb{Cl} or must use the complicated nonlinear version \rlb{new} (or \rlb{newnu}), depending on the paths from $\alpha$ to $\alpha'$}\/.
As far as we know, such an interesting (and a little annoying) ``twist'' has never been encountered in equilibrium physics.
It may be regarded as a sharp characterization of the difficulty of thermodynamics for NESS.

\section{Derivation}
In what follows we derive the extended Clausius relation \rlb{Cl} and its improvement \rlb{new}.
Although \rlb{Cl} was derived in \cite{1}, we here present a new derivation which sheds better light on the nature of the relation.

Since the derivation is somewhat involved, let us give some outline.
We follow the standard procedure to approximate a continuous protocol $\ah=(\alpha(t))_{t\in[0,\tf]}$ by a piecewise constant protocol which exhibits discontinuous jumps separated by a time interval of $2\tau$, where $\tau$ is a fixed time scale which is much larger than the relaxation time.
This procedure is carefully explained in section~\ref{s:steptoqs}.

When deriving the extended Clausius relations, we focus on the step protocol \rlb{stepa}, which exhibits a single jump within the time interval $[-\tau,\tau]$.
This corresponds to a part of the piecewise constant protocol mentioned above.
One of the keys to the derivation is the expression \rlb{exEP} of the excess entropy production in this step protocol.
The expression  \rlb{exEP} only involves the difference of the probability distributions of NESS with different parameters, and the entropy production $\sbkt{\Theta_{(\alpha')}}^{(\alpha')}_{\Gs}$ in a path starting from $\Ga$.

The most important ingredient in our derivation is the representation \rlb{KN}, which expresses the probability distribution $\rsa(\Gamma)$ of NESS in terms of the entropy production.
By combining this representation (or the similar representation \rlb{LR} in the linear response regime) with the above mentioned expression \rlb{exEP2} of the excess entropy production as well as the definition \rlb{Sa} of our symmetrized entropy, we can derive the step-protocol-versions \rlb{Clstep} and \rlb{newstep} of the extended Clausius relation and its nonlinear nonequilibrium version, respectively.
To be slightly more precise, when deriving the nonlinear relation \rlb{new}, we have to carefully distinguish step protocols in which only the relative inverse temperatures vary and those in which only the overall parameters (i.e., the reference inverse temperature and the parameter $\nu$) vary.
This is again carefully explained in section~\ref{s:steptoqs}.

\subsection{Ingredients for the derivation}

\subsubsection{Some basic definitions}
\label{s:defs}
Throughout the present derivation, we fix a time $\tau>0$ which is much larger than  the relaxation time of the system.
We denote by $(\alpha)$ the special protocol in the time interval $[0,\tau]$ in which the parameters are fixed at a constant $\alpha=(\betaon;\nu)$.

As in \cite{KN,1,JSP}, we define partially constrained path averages of any function $f(\Gh)$ of the path $\Gh$ as
\eq
\sbkt{f}^{\ca}_{\Gs}:=\int\DGh\,\delta({\Ga(0)-\Ga})\,\calW_{\ca}(\Gh)\,f(\Gh)
\lb{fGs}
\en
and
\eq
\sbkt{f}^{\ca}_{\sG}:=\{\rsa(\Ga)\}^{-1}\int\DGh\,\rsa(\Ga(0))\,\calW_{\ca}(\Gh)\,\delta(\Ga(\tau)-\Ga)\,f(\Gh).
\lb{fsG}
\en
As the subscripts indicate, they represent averages over histories in which the initial state and the final state, respectively, are specified to be $\Ga$.

In the following derivations we consider a  step protocol $\ahs$ in the time interval $[-\tau,\tau]$ defined by 
\eq
\alpha_\mathrm{s}(t)=
\begin{cases}
\alpha=(\betaon;\nu) & t\in[-\tau,0) \\
\alpha'=(\beta_1',\ldots,\beta_n';\nu') & t\in[0,\tau]
\end{cases}
\lb{stepa}
\en
By $\Di$, we denote the dimensionless measure of the magnitude of the difference $\alpha'-\alpha$.
This is a special case of the $\Di$ introduced in the extended Clausius relation \rlb{Cl}.

\subsubsection{Representation of the excess entropy production}

A crucial observation for the derivations is that the excess entropy production in the protocol \rlb{stepa} has an explicit representation
\eq
\Teas=\iG\{\rsa(\Ga)-\rs_{\alpha'}(\Ga)\}\,\sbkt{\Theta_{(\alpha')}}^{(\alpha')}_{\Gs}\ ,
\lb{exEP}
\en
where $(\alpha')$  stands for the constant protocol in the time interval $[0,\tau]$.
To see this one notes that 
$\iG\rsa(\Ga)\,\sbkt{\Theta_{(\alpha')}}^{(\alpha')}_{\Gs}$
is the total entropy production in $t\in[0,\tau]$ (after the sudden change of the parameter from $\alpha$ to $\alpha'$), 
and $\iG\rs_{\alpha'}(\Ga)\,\sbkt{\Theta_{(\alpha')}}^{(\alpha')}_{\Gs}$ is the total entropy production in the steady state with $\alpha'$.
The Markovian nature of the dynamics is also essential here.

Note that the left-hand side of \rlb{exEP} is the quantity for the protocol $\ahs$ defined in the time interval $[-\tau,\tau]$ while the right-hand side involves the quantity for the protocol $(\alpha')$ in $[0,\tau]$.
The relation \rlb{exEP} is nevertheless valid since the excess entropy production in the time interval $[-\tau,0]$ is vanishing in the protocol $\ahs$.

It is convenient to define
\eq
\DrG=\rs_{\alpha'}(\Ga)-\rs_{\alpha}(\Ga),
\lb{DrG}
\en
which satisfies $\DrG=\odo$ and $\iG\DrG=0$.
Then \rlb{exEP} is written as
\eq
\Teas=-\iG\DrG\,\sbkt{\Theta_{(\alpha')}}^{(\alpha')}_{\Gs}.
\lb{exEP2}
\en

\subsubsection{From step protocol to a quasi-static protocol}
\label{s:steptoqs}
In the following derivations, we do not derive the extended Clausius relation \rlb{Cl} or its nonlinear improvement \rlb{new} directly, but prove them only for the step protocol \rlb{stepa}.
The extended Clausius relations \rlb{Cl} for the step protocol that we will derive in section~\ref{s:Cl} is
\eq
\Ss(\alpha')-\Ss(\alpha)=-\Teas+\oesd+\ods.
\lb{Clstep}
\en

To discuss the step protocol version of the new relation \rlb{new}, we first have to restrict the way the parameters are varied.
We consider the step protocol \rlb{stepa} with the restriction that
\eq
\beta'_k=\beta_k+\Dei\beta
\lb{bdb}
\en
for all $k=1,\ldots,n$ with a common (small) $\Dei\beta$.
We also choose the reference inverse temperature as $\beta'=\beta+\Dei\beta$.
The parameter $\nu$ of the Hamiltonian can vary without any restrictions.
For the step protocol \rlb{stepa} with this restriction, we will show in section~\ref{s:new} that 
\eq
\Ss(\alpha')-\Ss(\alpha)=-\Teas+\oh\bbkt{\Dei(\beta H)(\Ga(0));\Theta_{\ahs}}^{\ahs}
+\oecd+\ods,
\lb{newstep}
\en
where $\Dei(\beta H)(\Ga)$ is defined as 
\eq
\Dei(\beta H)(\Ga):=\beta'H_{\nu'}(\Ga)-\beta\,H_\nu(\Ga).
\lb{DH}
\en

To get the relations \rlb{Cl} and  \rlb{new} for a continuous quasi-static protocol $\ah$, we only need to sum up the relations \rlb{Clstep} and \rlb{newstep}.
This is a standard procedure, but let us explain the details since there is indeed a subtle point.

Let us fix an arbitrary continuous protocol $\ah_\mathrm{ref}=(\alpha_\mathrm{ref}(t))_{t\in[0,1]}$ as our reference, and write $\ai=\alpha_\mathrm{ref}(0)$ and $\af=\alpha_\mathrm{ref}(1)$.
Let $N$ be a positive integer, and consider the time interval $[0,\tf]$ with $\tf:=2N\tau$.
For simplicity we have set the initial time as $\ti=0$.
We define a piecewise-constant protocol $\ah$ on $[0,\tf]$ by
\eq
\alpha(t)=
\begin{cases}
\alpha_0&t\in[0,\tau]\\
\alpha_j&t\in[(2j-1)\tau,(2j+1)\tau)\ \text{for $j=1,\ldots,N-1$}\\
\alpha_N&t\in[(2N-1)\tau,2N\tau]
\end{cases}
\lb{apc}
\en
where the sequence $\alpha_0,\alpha_1,\ldots,\alpha_N$ is a discrete approximation to $\alpha_\mathrm{ref}(t)$.
For the moment one can naively set $\alpha_j=\alpha_\mathrm{ref}(j/N)$ (but see below).
We now decompose the whole time interval $[0,\tf]=[0,2N\tau]$ into $N$ intervals $[2(j-1)\tau,2j\tau]$ with $j=1,\ldots,N$.
Denote by $\ah_j$ the protocol $\ah$ restricted on $[2(j-1)\tau,2j\tau]$.
Note that $\ah_j$ is a step protocol where the parameters jump from $\alpha_{j-1}$ to $\alpha_j$ in the middle point $t=(2j-1)\tau$.
We can then identify each interval $[2(j-1)\tau,2j\tau]$ and the associated protocol $\ah_j$ with the  interval $[-\tau,\tau]$ and the step protocol $\ahs$ of \rlb{stepa}, respectively.

Now by using the extended Clausius relation \rlb{Clstep} for the step protocol for each $\ah_j$, we find
\eqa
\Ss(\af)-\Ss(\ai)&=\sum_{j=1}^N\biggl\{\Ss\Bigl(\alpha_\mathrm{ref}\Bigl(\frac{j}{N}\Bigr)\Bigr)-\Ss\Bigl(\alpha_\mathrm{ref}\Bigl(\frac{j-1}{N}\Bigr)\Bigr)\biggr\}
\nl
&=\sum_{j=1}^N\Bigl\{-\sbkt{\Thetae_{\ah_j}}^{\ah_j}+O(\epsilon^2\Di_j)+O((\Di_j)^2)\Bigr\}
\nl
&=-\sbkt{\Thetae_{\ah}}^{\ah}+\oesd+O\Bigl(\frac{\Di^2}{N}\Bigr),
\ena 
where $\Di_j$ denotes the amount of the change of parameters in $\ah_j$.
We noted that $\Thetae_{\ah}=\sum_{j=1}^N\Thetae_{\ah_j}$, and wrote $\Di=\sum_{j=1}^N\Di_j$, and noted that $\Di_j=O(\Di/N)$.
It is crucial to note that each $\sbkt{\Thetae_{\ah_j}}^{\ah_j}$ is a quantity of $O(\delta_j)$ and hence one can sum them up over the whole (long) time interval.
By letting $N\uparrow\infty$, which is the quasi-static limit, we recover the desired extended Clausius relation \rlb{Cl}.

To get the nonlinear improvement \rlb{new}, we need a further care.
When defining the piecewise constant protocol \rlb{apc}, we make small modifications so that  the small step protocol $\ah_j$ satisfies the following.
In the protocol $\ah_j$ with an even $j$, the parameter $\nu$ do not vary.
The inverse temperatures $\beta_k$ change but with a restriction that the reference inverse temperature $\beta$ defined by \rlb{bref} do not change.
In the protocol $\ah_j$ with an odd $j$, on the other hand, all the inverse temperatures vary at once so that to satisfy \rlb{bdb} for all $k=1,\ldots,n$ (see section~\ref{s:new}).
The parameter $\nu$ may vary without any restrictions.

Then for $\ah_j$ with an even $j$, we use the extended Clausius relation \rlb{Clstep}, and for $\ah_j$ with an odd $j$, we use the improved relation \rlb{newstep}.
Again by summing all the relations, we have
\eqa
\Ss(\af)-&\Ss(\ai)=\sum_{j=1}^N\biggl\{\Ss\Bigl(\alpha_\mathrm{ref}\Bigl(\frac{j}{N}\Bigr)\Bigr)-\Ss\Bigl(\alpha_\mathrm{ref}\Bigl(\frac{j-1}{N}\Bigr)\Bigr)\biggr\}
\nl
&=\sum_{j:\,\mathrm{even}}\Bigl\{-\sbkt{\Thetae_{\ah_j}}^{\ah_j}+O(\epsilon^2\Di_j)+O((\Di_j)^2)\Bigr\}
\nl
&+\sum_{j:\,\mathrm{odd}}\Bigl\{-\sbkt{\Thetae_{\ah_j}}^{\ah_j}+\oh\bbkt{\Dei(\beta H)_j;\Theta_{\ah_j}}^{\ah_j}+O(\epsilon^3\Di_j)+O((\Di_j)^2)\Bigr\}
\nl
&=\sum_{j=1}^N\Bigl\{-\sbkt{\Thetae_{\ah_j}}^{\ah_j}+\oh\bbkt{\Dei(\beta H)_j;\Theta_{\ah_j}}^{\ah_j}\Bigr\}
+O(\epsilon^2\Di_\mathrm{r})+O(\epsilon^3\Di_\mathrm{o})+O\Bigl(\frac{\Di^2}{N}\Bigr),
\ena
where we wrote $\Di_\mathrm{r}=\sum_{j:\,\mathrm{even}}\Di_j$, $\Di_\mathrm{o}=\sum_{j:\,\mathrm{odd}}\Di_j$ and $\Di=\sum_{j=1}^N\Di_j$.
We also noted that
\eq
\Dei(\beta H)_j:=\beta\bigl(2(j-1)\tau\bigr)\,H_{\nu_j}\Bigl(\Ga\bigl((2j-1)\tau\bigr)\Bigr)-\beta\bigl(2(j+1)\tau\bigr)\,H_{\nu_{j-1}}\Bigl(\Ga\bigl((2j-1)\tau\bigr)\Bigr)
\lb{DHj}
\en
vanishes for an even $j$, where $\beta(t)$ denotes the reference inverse temperature at time $t$.
Since $\tau$ is chosen to be much larger than the relaxation time, the correlation between $\Dei(\beta H)_j$ and $\Theta_{\ah_j}$ can be safely neglected if $j\ne j'$.
We can then sum up the result to get
\eq
\Ss(\af)-\Ss(\ai)=-\sbkt{\Thetae_{\ah}}^{\ah}+\frac{1}{2}\Bbkt{\sum_{j=1}^N\Dei(\beta H)_j;\Theta_{\ah}}^{\ah}+O(\epsilon^2\Di_\mathrm{r})+O(\epsilon^3\Di_\mathrm{o})+O\Bigl(\frac{\Di^2}{N}\Bigr).
\lb{SSnewp}
\en
Since these quantities vary every $4\tau$ in the discrete approximation, we can write
\eq
\Dei(\beta H)_j= 4\tau\,\frac{d}{ds}\{\beta(s)\,H_{\nu(s)}\}\Bigr|_{s=(2j-1)\tau}+O\Bigl(\frac{1}{N^2}\Bigr),
\en
where $\beta(s)$ and $\nu(s)$ in the right-hand side denote (with a slight abuse of notation) the smooth approximations of the reference inverse temperature and the parameter, respectively.
Since this means that
\eq
\sum_{j=1}^N\Dei(\beta H)_j=\int_0^{\tf}dt\,\frac{d}{ds}\{\beta(s)\,H_{\nu(s)}\}\Bigr|_{s=t}
+O\Bigl(\frac{1}{N}\Bigr).
\en
we get the desired \rlb{new} by letting $N\uparrow\infty$ in \rlb{SSnewp}.

\subsubsection{Representation of the probability distribution of NESS}
The most important ingredient of our derivation is the following representation for the probability distribution of NESS derived by two of us in \cite{KN}
\eq
\log\rsa(\Ga)=-\tilde{S}(\alpha)
+\frac{1}{2}\bigl\{\sbkt{\Theta_{\ca}}^{\ca}_{\sG}
-\sbkt{\Theta_{\ca}}^{\ca}_{\Gss}\bigr\}+\oec,
\lb{KN}
\en
where $(\alpha)$ is the constant protocol in $[0,\tau]$ defined in section~\ref{s:defs}, and  $\tilde{S}(\alpha)$ is a normalization constant.
We can indeed show that $\tilde{S}(\alpha)=\Ss(\alpha)+\oec$, but this fact is not used in the present derivation.  See \cite{1,2}.

Let us use the reference temperature $\beta$ of \rlb{bref}, and define the ``nonequilibrium part'' of the entropy production as
\eqa
\Psi_{\ca}(\Gh)&:=-\int_{0}^{\tau}dt\sumn(\beta_k-\beta)\,J_k(\Gh;t)
\nl
&=\Theta_{\ca}(\Gh)+\beta\{H_{\nu}(\Gamma(\tau))-H_{\nu}(\Gamma(0))\}.
\lb{Psi}
\ena
To get the second line, we noticed that $\sum_{k=1}^nJ_k(\Gh;t)$ is the total heat current that flows into the system, and used the energy conservation.
With \rlb{Psi}, the representation \rlb{KN} is rewritten as
\eqa
\log\rsa(\Ga)&=\beta\,\tilde{F}(\alpha)-\beta\,H_\nu(\Gamma)
+\frac{1}{2}\bigl\{\sbkt{\Psi_{\ca}}^{\ca}_{\sG}
-\sbkt{\Psi_{\ca}}^{\ca}_{\Gss}\bigr\}+\oec
\nl
&=\beta\,\tilde{F}(\alpha)-\beta\,H_\nu(\Gamma)
-\psi_\alpha(\Ga)+\oec,
\lb{KN2}
\ena
where we defined another constant (free energy) $\tilde{F}(\alpha):=\int d\Ga\,\rsa(\Ga)\,H_\nu(\Ga)-\tilde{S}(\alpha)/\beta$.
We have introduced a new quantity
\eq
\psi_\alpha(\Ga):=\frac{1}{2}\{\sbkt{\Psi_{\ca}}^{\ca}_{\Gss}-\sbkt{\Psi_{\ca}}^{\ca}_{\sG}\},
\lb{psi}
\en
which represents the nonequilibrium correction to the canonical distribution.
Note that $\Psi_{\ca}(\Gh)$ is typically a quantity of $\oeo$, and so is $\psi_\alpha(\Ga)$.

Now for a general function $f(\Gh)$ of the path $\Gh$, a naive perturbation around the equilibrium implies
\eq
\sbkt{f}^{\ca}_{\sG}=\sbkt{f}^{\caeq}_{\mathrm{eq}\to\Ga}+O(f)\,\oeo,
\en
where for $\alpha=(\betaon;\nu)$ we set $\alphaeq=(\beta,\ldots,\beta;\nu)$ with $\beta$ of \rlb{bref}.
By setting $f=\Psi_{\ca}$, we have
\eq
\sbkt{\Psi_{\ca}}^{\ca}_{\sG}=\sbkt{\Psi_{\ca}}^{\caeq}_{\mathrm{eq}\to\Ga}+\oes,
\lb{Psiper}
\en
because $O(\Psi_{\ca})=\oeo$.
It is crucial to note that the function in the expectation value in the right-hand side is $\Psi_{\ca}(\Gh)$, rather than $\Psi_{\caeq}(\Gh)$.

Since the equilibrium path average has an exact time-reversal symmetry\footnote{
This is a standard result (see, for example, \cite{JSP} for details).
For a path $\Gh=(\Ga(t))_{t\in[0,\tau]}$ we define its time reversal as $\Gh^\dagger=((\Ga(\tau-t))^*)_{t\in[0,\tau]}$.
Then for a function $f(\Gh)$ its time reversal is defined by $f^\dagger(\Gh)=f(\Gh^\dagger)$.
Then the equilibrium dynamics satisfies
$\sbkt{f}^{\caeq}_{\mathrm{eq}\to\Ga}=\sbkt{f^\dagger}^{\caeq}_{\Gas\to\mathrm{eq}}$.
Finally noting that $J_k(\Gh^\dagger;\tau-t)=-J_k(\Gh;t)$, \rlb{Psi} implies $(\Psi_{\ca})^\dagger=-\Psi_{\ca}$.
}
$\sbkt{\Psi_{\ca}}^{\caeq}_{\mathrm{eq}\to\Ga}=-\sbkt{\Psi_{\ca}}^{\caeq}_{\Gas\to\mathrm{eq}}$, 
we further have
$\sbkt{\Psi_{\ca}}^{\ca}_{\sG}=-\sbkt{\Psi_{\ca}}^{\caeq}_{\Gas\to\mathrm{eq}}+\oes
=-\sbkt{\Psi_{\ca}}^{\ca}_{\Gss}+\oes$,
where we used the relation for $\sbkt{\Psi_{\ca}}^{\ca}_{\Gss}$ corresponding to \rlb{Psiper}.
We have thus shown the symmetry
\eq
\sbkt{\Psi_{\ca}}^{\ca}_{\sG}+\sbkt{\Psi_{\ca}}^{\ca}_{\Gss}=\oes,
\lb{sym}
\en
which will be useful.

By substituting the symmetry \rlb{sym} into the representation \rlb{KN2}, we see
\eqa
\log\rsa(\Ga)&=\beta\,\tilde{F}(\alpha)-\beta\,H_\nu(\Gamma)
-\sbkt{\Psi_{\ca}}^{\ca}_{\Gss}+\oes
\nl
&=-\tilde{S}(\alpha)-\sbkt{\Theta_{\ca}}^{\ca}_{\Gss}+\oes,
\lb{LR}
\ena
which indeed is a slightly different way of expressing the standard linear response formula.
See the remark below.
We shall see in section~\ref{s:Cl} that this representation is sufficient to derive the (older) extended Clausius relation \rlb{Cl}.

Let us note that the terms $\sbkt{\Psi_{\ca}}^{\ca}_{\sG}$ and 
$\sbkt{\Psi_{\ca}}^{\ca}_{\Gss}$ in \rlb{sym} and $\sbkt{\Theta_{\ca}}^{\ca}_{\Gss}$ in \rlb{LR} all contain a contribution which grows proportionally with $\tau$, namely, the total entropy production $\sigma(\alpha)\,\tau$.
This means that the error terms $\oes$ in \rlb{sym} and \rlb{LR} include the same $\tau$ linear contribution.
Fortunately this $\tau$ linear contribution is {\em not}\/ inherited by the error terms in the extended Clausius relation \rlb{Cl} or its improvement \rlb{new}.
This is because only $\Gamma$ dependent quantities enter into these final relations, and the $\tau$ linear contribution, which is $\sigma(\alpha)\,\tau$, just drops out.
See the derivation for details.

\bn
{\em Remark:}\/
Let us make a few remarks about the representations of the probability distribution of NESS that we have discussed.

As we have noted above, the term $\sbkt{\Theta_{\ca}}^{\ca}_{\Gss}$ in the
representation \rlb{LR} contains the total entropy production $\sigma(\alpha)\,\tau$, which grows linearly in  $\tau$.
This is not quite desirable since one usually uses this kind of representation
in the limit $\tau\uparrow\infty$.

One natural way to get rid of the $\tau$ linear divergence is to define the excess quantity corresponding to \rlb{Psi} by
\eqa
\Psi^\mathrm{ex}_{\ca}(\Gh)&:=-\int_{0}^{\tau}dt\sumn(\beta_k-\beta)\,\{J_k(\Gh;t)-J^\mathrm{ss}_k(\alpha)\}
\nl
&=\Psi_{\ca}(\Ga)+\tau\,\sum_{k=1}^n(\beta_k-\beta)\,J^\mathrm{ss}_k(\alpha)=\Psi_{\ca}(\Ga)+\oes,
\lb{Psiex}
\ena
where $J^\mathrm{ss}_k(\alpha)=\oeo$ is the steady heat current.
We can then readily rewrite \rlb{LR} as
\eq
\log\rsa(\Ga)=\beta\,\tilde{F}(\alpha)-\beta\,H_\nu(\Gamma)
-\sbkt{\Psi^\mathrm{ex}_{\ca}}^{\ca}_{\Gss}+\oes.
\en
In this new expression $\sbkt{\Psi^\mathrm{ex}_{\ca}}^{\ca}_{\Gss}$ no longer contains a term proportional to $\tau$.
One can safely take the  $\tau\uparrow\infty$ limit here\footnote{
Essentially the same consideration applies to the symmetry relation \rlb{sym} since the left-hand side of \rlb{sym} has a contribution proportional to $\tau$.
By using \rlb{Psiex}, \rlb{sym} implies $\sbkt{\Psi^\mathrm{ex}_{\ca}}^{\ca}_{\sG}+\sbkt{\Psi^\mathrm{ex}_{\ca}}^{\ca}_{\Gss}=\oes$.
In this form, the left-hand side does not contain any term which is proportional to $\tau$.
}.

Another natural cure, which is  suggested by \rlb{Psiper} and \rlb{sym}, is to rewrite the representation \rlb{LR} as 
\eq
\log\rsa(\Ga)=\beta\,\tilde{F}(\alpha)-\beta\,H_\nu(\Gamma)
-\sbkt{\Psi_{\ca}}^{\caeq}_{\Gas\to\mathrm{eq}}+\oes,
\lb{LR2}
\en
which is a standard linear response formula for the probability distribution of NESS.
Note that one can safely take the $\tau\uparrow\infty$ limit \rlb{LR2} since the entropy production rate in the  equilibrium vanishes.

It should be stressed in passing that the higher order representations \rlb{KN}, \rlb{KN2} do not suffer from the problem of $\tau$ dependence.
Since both $\sbkt{\Theta_{\ca}}^{\ca}_{\sG}$ and $\sbkt{\Theta_{\ca}}^{\ca}_{\Gss}$ are proportional to $\sigma(\alpha)\,\tau$ for large $\tau$, the divergence nicely cancels out.
Thus one can also take the $\tau\uparrow\infty$ limit in the representations \rlb{KN}, \rlb{KN2}.

\subsection{Derivation of the extended Clausius relation \rlb{Clstep} --- easier case}
In what follows, we abbreviate $\rsa$, $\rs_{\alpha'}$, $\Theta_{\ca}$, $\Theta_{(\alpha')}$, $\Psi_{\ca}$, $\Psi_{(\alpha')}$, $\sbkt{\cdots}^{\ca}$, and $\sbkt{\cdots}^{(\alpha')}$ as
$\rho$, $\rho'$, $\Theta$, $\Theta'$, $\Psi$, $\Psi'$, $\sbkt{\cdots}$, and $\sbkt{\cdots}'$, respectively.

As a warm up, we shall derive the extended Clausius relation \rlb{Clstep} in a restricted class of models where one has $\rsa(\Ga)=\rsa(\Gas)$, i.e., the probability distribution in NESS has a time-reversal symmetry\footnote{
Over damped models without momenta posses this symmetry.
}.
Although such a symmetry is in general absent, the derivation clarifies the relation between the linear response formula \rlb{LR} and the extended Clausius relation \rlb{Clstep}.
We consider the step protocol \rlb{stepa} where arbitrary change from $\alpha$ to $\alpha'$ is allowed.

In this case the entropy \rlb{Sa} coincides with the Shannon entropy
\eq
S_\mathrm{Sh}(\alpha):=-\iG\rsa(\Ga)\,\log\rsa(\Ga).
\lb{Sh}
\en
From \rlb{DrG} and a Taylor expansion, one has
\eqa
S_\mathrm{Sh}(\alpha)&=-\iG\{\rpG-\DrG\}\,\log\{\rpG-\DrG\}
\nl
&=-\iG\rpG\,\log\rpG+\iG\DrG\,\log\rpG-\iG\rpG\frac{\DrG}{\rpG}+\ods
\nl
&=S_\mathrm{Sh}(\alpha')+\iG\DrG\,\log\rpG+\ods
\ena
Thus by using the assumption $\rpG=\rpGs$ we see that
\eqa
S_\mathrm{Sh}(\alpha')-S_\mathrm{Sh}(\alpha)&=-\iG\DrG\,\log\rpGs+\ods
\nl
&=\iG\DrG\,\sbkt{\Theta_{\cpa}}^{\cpa}_{\Gs}+\oesd+\ods,
\lb{ShTheta}
\ena
where we used \rlb{LR} and noted that $\iG\DrG\,{\rm const.}=0$.
As we have explained just below \rlb{LR}, all terms which are independent of $\Gamma$ (including the total entropy production $\sigma(\alpha)\,\tau$) disappears in this process.
From the representation \rlb{exEP2} we get the desired extended Clausius relation
\eq
S_\mathrm{Sh}(\alpha')-S_\mathrm{Sh}(\alpha)=-\Teas+\oesd+\ods
\lb{ShTheta2}
\en
for the step protocol of \rlb{stepa}.

Let us also note that the above manipulation in general case (without the assumption $\rsa(\Ga)=\rsa(\Gas)$) leads us to the expression
\eq
S_\mathrm{Sh}(\alpha')-S_\mathrm{Sh}(\alpha)=\iG\DrG\,\sbkt{\Theta_{\cpa}}^{\cpa}_{\Gss}+\oesd+\ods.
\en
The quantity in the right-hand side can be measured numerically (but not experimentally).
This expression was used in \cite{KNI} for a numerical evaluation of the Shannon entropy in a NESS.

\subsection{Derivation of the extended Clausius relation \rlb{Clstep}}
\label{s:Cl}
Now let us turn to the derivation of the extended Clausius relation \rlb{Clstep} in the general case.
Here we shall make use of the property
\eq
\rG-\rGs=\oeo,
\lb{rho-rho}
\en
which follows from \rlb{LR}.
We again consider the step protocol \rlb{stepa} where arbitrary change from $\alpha$ to $\alpha'$ is allowed.
 
From the definition \rlb{Sa} of our symmetrized entropy, we find
\eqa
&\Ss(\alpha')-\Ss(\alpha)
\nl
&=-\oh\iG\{\rG+\DrG\}\{\log\rpG+\log\rpGs\}
+\oh\iG\rG\{\log\rG+\log\rGs\}
\nl
&=-\iG\DrG\,\log\rpGs+\oh\iG\DrG\{\log\rpGs-\log\rpG\}
\nl
&\hspace{0.5cm}-\oh\iG\rG\{\log\rpG-\log\rG\}-\oh\iG\rG\{\log\rpGs-\log\rGs\}
\nl
&=:A_1+A_2+A_3+A_4,
\lb{SSA}
\ena
where the definitions of $A_1$, $A_2$, $A_3$, and $A_4$ should be evident.

Let us evaluate these terms separately.
As for $A_1$, we simply repeat the manipulation in \rlb{ShTheta}, \rlb{ShTheta2} to conclude
\eq
A_1=-\Teas+\oesd.
\en
As for $A_2$, we make use of the expansion in $\rG-\rGs=\oeo$ as
\eq
\log\rGs=\log\bigl[\rG+\{\rGs-\rG\}\bigr]
=\log\rG+\frac{\rGs-\rG}{\rG}+\oes
\en
to find that
\eqa
A_2&=\oh\iG\DrG\{\log\rGs-\log\rG\}+\ods
\nl
&=\oh\iG\DrG\,\Bigl\{\frac{\rGs}{\rG}-1+\oes\Bigr\}+\ods
\nl
&=\oh\iG\DrG\frac{\rGs}{\rG}+\oesd+\ods.
\ena
The treatment of $A_3$ is easy.
By a simple expansion, we see that
\eqa
A_3&=-\oh\iG\rG\,\bigl[\log\{\rG+\DrG\}-\log\rG\bigr]
\nl
&=-\oh\iG\rG\,\Bigl\{\frac{\DrG}{\rG}+\ods\Bigr\}
\nl
&=-\oh\iG\DrG+\ods=\ods.
\lb{A3}
\ena
As for $A_4$, we repeat this estimate to get
\eq
A_4=-\oh\iG\rG\,\Bigl\{\frac{\DrGs}{\rGs}+\ods\Bigr\}
=-\oh\iG\rGs\,\frac{\DrG}{\rG}+\ods,
\en
where made a change of variable $\Ga\to\Gas$.

By summing up the results, $A_2$ and $A_4$ (rather surprisingly) cancel out, and we get the desired
\eq
\Ss(\alpha')-\Ss(\alpha)=-\Teas+\oesd+\ods, 
\en
for the step protocol of \rlb{stepa}.

\subsection{Derivation of the improved relation \rlb{newstep}}
\label{s:new}
Now let us turn to the derivation of the ``nonlinear'' relation \rlb{newstep}.

Unlike in the previous two subsections, we only consider the step protocol  \rlb{stepa} which satisfies the condition \rlb{bdb} that all the inverse temperatures $\beta_1,\ldots,\beta_n$ of the baths and the reference inverse temperature $\beta$ shift by the same (small) amount.
This means that we have
\eq
\beta_k-\beta=\beta'_k-\beta'
\lb{bbk}
\en
for all $k=1,\ldots,n$.
The parameter $\nu$ of the Hamiltonian can vary without any restrictions.

When (and only when) we make this restriction, we have
\eq
\DpG:=\psi_{\alpha'}(\Ga)-\psi_\alpha(\Ga)=\oeod.
\lb{DpG}
\en
To see this one notes from the definition \rlb{Psi} and the condition \rlb{bbk}
\eq
\bpsG{\Psi'}-\bsG{\Psi}=\sum_{k=1}^n(\beta_k-\beta)
\biggl\{
\Bbkt{\int_0^\tau dt\,J_k(\Gh;t)}_{\sG}'-\Bbkt{\int_0^\tau dt\,J_k(\Gh;t)}_{\sG}
\biggr\}
=\oeod
\lb{PpP}
\en
because $\beta_k-\beta=\oeo$ and $\bpsG{\cdots}-\bsG{\cdots}=\odo$.
With the corresponding estimate for $\bpGss{\Psi'}-\bGss{\Psi}$ and the definition \rlb{psi}, we get\footnote{
Note that $\psi_\alpha(\Ga)$ defined by \rlb{psi} contains no terms which are proportional to $\tau$, while both
$\sbkt{\Psi_{\ca}}^{\ca}_{\Gss}$ and $\sbkt{\Psi_{\ca}}^{\ca}_{\sG}$
contain the total entropy production $\sigma(\alpha)\,\tau$.
Thus \rlb{PpP} has a term of $\oes$ which is proportional to $\tau$, but the desired \rlb{DpG} has no such terms.
} \rlb{DpG}.

We also use the quantity
\eqa
\deltan(\Ga)&:=\frac{\DrG}{\rG}=\log\rpG-\log\rG+\ods
\nl
&=
\{\beta'\tilde{F}(\alpha')-\beta'H_{\nu'}(\Ga)-\psi_{\alpha'}(\Ga)\}
-\{\beta\,\tilde{F}(\alpha)-\beta\,H_\nu(\Ga)-\psi_\alpha(\Ga)\}
+\oecd+\ods
\nl
&=:
\Dei(\beta\tilde{F})-\Dei(\beta H)(\Ga)-\DpG+\oecd+\ods,
\lb{delta}
\ena
where we used \rlb{KN2} noting that\footnote{
To see this one expands $\log\rG$ into a power series in $n$ quantities $\beta_k-\beta$ ($k=1,\ldots,n$) where the coefficients are functions of $\beta$, $\nu$, and $\Ga$.
Since $\beta_k-\beta$ are the same for $\log\rG$ and $\log\rpG$ and only the coefficients change by $\odo$, we see that the difference of the $\oec$ terms is $\oecd$.
} the difference of the terms of $\oec$ is $\oecd$ for the present protocol with the restriction \rlb{bbk}.
The final line defines $\Dei(\beta\tilde{F})$ and $\Dei(\beta H)(\Ga)$ (see also \rlb{DH}).

Note that 
\eq
\deltan(\Ga)-\deltan(\Gas)=-\DpG+\Dei\psi(\Gas)+\oecd+\ods=\oeod+\ods.
\lb{dd}
\en
This implies the useful fact
\eqa
\Dei\rho_-(\Ga)&:=\oh\{\DrG-\DrGs\}
\nl&
=\oh\{\rG\,\deltan(\Ga)-\rGs\,\deltan(\Gas)\}
\nl
&=\frac{1}{4}\{\rG-\rGs\}\{\deltan(\Ga)+\deltan(\Gas)\}
+\frac{1}{4}\{\rG+\rGs\}\{\deltan(\Ga)-\deltan(\Gas)\}
\nl
&=\oeod+\ods,
\lb{Drm}
\ena
where we noted that $\rG-\rGs=\oeo$, $\deltan(\Ga)+\deltan(\Gas)=\odo$, and \rlb{dd}.

From the definition \rlb{Sa} of our symmetrized entropy, we observe that
\eqa
&\Ss(\alpha')-\Ss(\alpha)
\nl
&=-\oh\iG\{\rG+\DrG\}\{\log\rpG+\log\rpGs\}
+\oh\iG\rG\{\log\rG+\log\rGs\}
\nl
&=-\oh\iG\DrG\{\log\rpG+\log\rpGs\}-\iG\rG\{\log\rpG-\log\rG\}
\nl
&\hspace{0.4cm}+\oh\iG\rG\{\log\rpG-\log\rpGs\}-\oh\iG\rG\{\log\rG-\log\rGs\}
\nl
&=:B_1+B_2+B_3+B_4,
\lb{SSB}
\ena
where the definitions of $B_1$, $B_2$, $B_3$, and $B_4$ should be evident.
$B_2$ is the same as $A_3$ in \rlb{SSA} and hence $B_2=\ods$ from \rlb{A3}.
We shall evaluate $B_1$ and $B_3+B_4$.

We shall now substitute the higher order representation \rlb{KN} of $\rsa(\Ga)$ into the definition of $B_1$.
Noting that $\iG\DrG\{\mathrm{const.}\}=0$, we organize $B_1$ as
\eqa
B_1&=-\frac{1}{4}\iG\DrG\,
\bigl\{\bpsG{\Theta'}-\bpGss{\Theta'}+\bpsGs{\Theta'}-\bpGs{\Theta'}\bigr\}+\oecd
\nl
&=\iG\DrG\,\bpGs{\Theta'}-\oh\iG\DrG\,\bigl\{\bpsG{\Theta'}+\bpGs{\Theta'}\bigr\}
\nl
&\hspace{0.4cm}+\frac{1}{4}\iG\DrG\,
\bigl\{\bpsG{\Theta'}+\bpGss{\Theta'}-\bpsGs{\Theta'}-\bpGs{\Theta'}\bigr\}+\oecd
\nl&=:C_1+C_2+C_3+\oecd.
\lb{B1}
\ena
Note that $C_1$ is indeed what we want because \rlb{exEP2} reads
\eq
C_1=-\Teas.
\lb{C1}
\en
Let us rewrite $C_2$ as
\eq
C_2=-\oh\iG\DrG\,\bigl\{\bsG{\Theta}+\bpGs{\Theta'}\bigr\}+\ods
=-\oh[\deltan(\Ga);G(\Ga)]_\rho+\ods,
\lb{C2}
\en
where we used $\bpsG{\Theta'}=\bsG{\Theta}+\odo$, and defined
\eq
G(\Ga):=\bsG{\Theta}+\bpGs{\Theta'}
\lb{GG}
\en
We also introduced the average
\eq
[A(\Ga)]_\rho:=\iG\rG\,A(\Ga)
\lb{[A]}
\en
and the corresponding truncated average
\eq
[A(\Ga);B(\Ga)]_\rho:=[A(\Ga)\,B(\Ga)]_\rho-[A(\Ga)]_\rho\,[B(\Ga)]_\rho
\en
for any functions $A(\Ga)$ and $B(\Ga)$ of $\Ga$.
In \rlb{C2}, one first gets $[\deltan(\Ga)\,G(\Ga)]_\rho$, but this can be replaced by  the truncated average since $[\deltan(\Ga)]_\rho=\iG\DrG=0$.

To deal with $C_3$, we note that $\bigl\{\bpsG{\Theta'}+\bpGss{\Theta'}-\bpsGs{\Theta'}-\bpGs{\Theta'}\bigr\}$ changes the sign when we replace $\Ga$ with $\Gas$.
We therefore recall \rlb{Drm} and write
\eqa
C_3&=\frac{1}{4}\iG\Dei\rho_-(\Ga)\,
\bigl\{\bpsG{\Theta'}+\bpGss{\Theta'}-\bpsGs{\Theta'}-\bpGs{\Theta'}\bigr\}
\nl
&=\frac{1}{4}\iG\Dei\rho_-(\Ga)\,
\bigl\{\bpsG{\Psi'}+\bpGss{\Psi'}-\bpsGs{\Psi'}-\bpGs{\Psi'}\bigr\},
\lb{C3p}
\ena
where we used \rlb{Psi}.
Now recalling the approximate time-reversal symmetry \rlb{sym}, one finds\footnote{
Note that $\tau$ linear contributions completely cancel out in the sum.
} $\{\cdots\}=\oes$.
Combining this estimate with \rlb{Drm}, we see that
\eq
C_3=\oecd+O(\epsilon^2\Di^2).
\lb{C3}
\en

It now remains to evaluate $B_3+B_4$.
By making a change of variable $\Ga\to\Gas$ in $B_4$, we can rewrite $B_3+B_4$ as
\eqa
B_3+B_4&=-\oh\iG\{\rGs-\rG\}\{\log\rpG-\log\rG\}
\nl
&=-\oh\iG\{\rGs-\rG\}\{\Dei(\beta\tilde{F})-\Dei(\beta H)(\Ga)-\DpG\}
+O(\epsilon^4\Di)+O(\epsilon\Di^2)
\itext{where we used \rlb{delta}.
Noting that $\Dei(\beta\tilde{F})-\Dei(\beta H)(\Ga)$ is invariant under $\Ga\to\Gas$, we get}
&=\oh\iG\{\rGs-\rG\}\,\DpG
+O(\epsilon^4\Di)+O(\epsilon\Di^2)
\nl
&=\oh\biggl[\Bigl(\frac{\rGs}{\rG}-1\Bigr);\DpG\biggr]_\rho+O(\epsilon^4\Di)+O(\epsilon\Di^2).
\lb{B34p}
\ena
Here we used the truncated average since
\eq
\biggl[\frac{\rGs}{\rG}-1\biggr]_\rho=\iG\{\rGs-\rG\}=0.
\en
By using \rlb{KN2}, and recalling that $H_\nu(\Gas)=H_\nu(\Ga)$, we see
\eq
\frac{\rGs}{\rG}-1=\exp\bigl[-\psi(\Gas)+\psi(\Ga)+\oec\bigr]-1
=\psi(\Ga)-\psi(\Gas)+\oes
\en
because $\psi(\Ga)=\oeo$.
We substitute this into \rlb{B34p} and recall that $\DpG=\oeod$ as in \rlb{DpG} to get
\eqa
B_3+B_4&=
\oh\bigl[\DpG;\{\psi(\Ga)-\psi(\Gas)\}\bigr]_\rho+\oecd+O(\epsilon\Di^2)
\nl
&=\frac{1}{4}\Bigl[\DpG;\bigl\{\bGss{\Psi}-\bsG{\Psi}-\bGs{\Psi}+\bsGs{\Psi}\bigr\}\Bigr]_\rho+\oecd+O(\epsilon\Di^2)
\itext{where we used the definition \rlb{psi} of $\psi(\Gamma)$.
Again from the approximate time-reversal symmetry \rlb{sym} and $\DpG=\oeod$, we have
}
&=-\oh\Bigl[\DpG;\bigl\{\bGs{\Psi}+\bsG{\Psi}\bigr\}\Bigr]_\rho+\oecd+O(\epsilon\Di^2)
\nl
&=-\oh\Bigl[\DpG;\bigl\{\bGs{\Theta}+\bsG{\Theta}\bigr\}\Bigr]_\rho+\oecd+O(\epsilon\Di^2)
\nl
&=-\oh[\DpG;G(\Ga)]_\rho+\oecd+\ods,
\lb{B34}
\ena
where we again noted that $\bpsG{\Theta'}=\bsG{\Theta}+\odo$, and used the definition \rlb{GG}.
Note that the $\Gamma$ independent $\tau$ linear contribution again plays no roles since $[f(\Ga);\text{const.}]=0$ for any function $f(\Ga)$.

Now by summing \rlb{C1}, \rlb{C2}, \rlb{C3}, and \rlb{B34}, we get
\eq
\Ss(\alpha')-\Ss(\alpha)=-\Teas-\oh\bigl[\{\deltan(\Ga)+\DpG\};G(\Ga)\bigr]_\rho
+\oecd+\ods.
\lb{SSdd}
\en
Since \rlb{delta} implies
\eq
\deltan(\Ga)+\DpG=\Dei(\beta\tilde{F})-\Dei(\beta H)(\Ga)+\oecd+\ods,
\en
\rlb{SSdd} is rewritten as
\eq
\Ss(\alpha')-\Ss(\alpha)=-\Teas+\oh\bigl[\Dei(\beta H)(\Ga);G(\Ga)\bigr]_\rho
+\oecd+\ods,
\lb{SSdd2}
\en
where $\Dei(\beta H)(\Ga)$ is defined in \rlb{DH}.

It now remains to interpret the truncated correlation in \rlb{SSdd2}.
We claim that, for any function $A(\Ga)$ of $\Ga$, one can rewrite the correlation function as
\eq
[A(\Ga)\,G(\Ga)]_\rho=\sbkt{A(\Ga(0))\,\Theta_{\ahs}}^{\ahs}
\lb{AG}
\en
where the right-hand side is the path average in the step protocol \rlb{stepa}.
By using this, \rlb{SSdd2} becomes
\eq
\Ss(\alpha')-\Ss(\alpha)=-\Teas+\oh\bbkt{\Dei(\beta H)(\Ga(0));\Theta_{\ahs}}^{\ahs}
+\oecd+\ods,
\lb{newstep2}
\en
which is our goal \rlb{newstep} in the present section.

To show the claim \rlb{AG}, we substitute the definitions \rlb{GG} and \rlb{[A]} to get
\eq
[A(\Ga)\,G(\Ga)]_\rho=\iG\rsa(\Ga)\,A(\Ga)\,\sbkt{\Theta_{\ca}}^{\ca}_{\sG}
+\iG\rsa(\Ga)\,A(\Ga)\,\sbkt{\Theta_{\cpa}}^{\cpa}_{\Gs}
\lb{AG2}
\en 
The key observation is to consider the step protocol \rlb{stepa}, and decompose a path $\Gh=(\Ga(t))_{t\in[-\tau,\tau]}$ in the interval $[-\tau,\tau]$ into two parts $\Ghm=(\Ga_-(t))_{t\in[-\tau,0]}$ and $\Ghp=(\Ga_+(t))_{t\in[0,\tau]}$.
One then identifies $\sbkt{\cdots}^{\ca}_{\sG}$ with the path average over $\Ghm$, and $\sbkt{\cdots}^{\cpa}_{\Gs}$ with that over $\Ghp$.
Then the claim \rlb{AG} may be intuitively clear, but let us present a careful derivation for completeness.

We first note that, since the process is Markovian,  the original path average is related to the  decomposed path averages through 
\eq
\int\DGh(\cdots)=\int\DGh_-\int\DGh_+\,\delta\bigl(\Ga_-(0)-\Ga_+(0)\bigr)(\cdots)
\en
and
\eq
\calW_{\ahs}(\Gh)=\calW_{\ca}(\Ghm)\,\calW_{\cpa}(\Ghp).
\en
By substituting the definition \rlb{fsG}, we find
\eqa
\iG&\rsa(\Ga)\,A(\Ga)\,\sbkt{\Theta_{\ca}}^{\ca}_{\sG}
\nl
&=\iG A(\Ga)\int\DGh_-\,\rsa(\Ga(-\tau))\,\calW_{\ca}(\Ghm)\,\delta(\Ga_-(0)-\Ga)\,\Theta_{\ca}(\Ghm)
\nl
&=\int\DGh_-\,\rsa(\Ga(-\tau))\,\calW_{\ca}(\Ghm)\,A(\Ga_-(0))\,\Theta_{\ca}(\Ghm)
\nl
&=\int\DGh\,\rsa(\Ga(-\tau))\,\calW_{\ahs}(\Gh)\,A(\Ga(0))\,\Theta_{\ca}(\Ghm),
\lb{AG3}
\ena
where we used the normalization $\int\DGh_+\,\delta(\Ga_+(0)-\Ga)\,\calW_{\cpa}(\Ghp)=1$.
Similarly we substitute the definition \rlb{fGs} to get
\eqa
\iG&\rsa(\Ga)\,A(\Ga)\,\sbkt{\Theta_{\cpa}}^{\cpa}_{\Gs}
\nl
&=\iG\rsa(\Ga)\,A(\Ga)\int\DGh_+\,\delta(\Ga_+(0)-\Ga)\,\calW_{\cpa}(\Ghp)\,\Theta_{\cpa}(\Ghp)
\nl
&=\int\DGh_+\,\rsa(\Ga_+(0))\,\calW_{\cpa}(\Ghp)\,A(\Ga_+(0))\,\Theta_{\cpa}(\Ghp)
\nl
&=\int\DGh\,\rsa(\Ga(-\tau))\,\calW_{\ahs}(\Gh)\,A(\Ga(0))\,\Theta_{\cpa}(\Ghp),
\lb{AG4}
\ena
where we used the invariance $\rsa(\Ga)=\int\DGh_-\,\rsa(\Ga_-(-\tau))\,\calW_{\ca}(\Ghm)\,\delta(\Ga_-(0)-\Ga)$ of the NESS.
Adding \rlb{AG3} and \rlb{AG4}, noting that $\Theta_{\ahs}(\Gh)=\Theta_{\ca}(\Ghm)+\Theta_{\cpa}(\Ghp)$, and recalling the early definition \rlb{Fa}, we get the desired claim \rlb{AG}.

Finally let us show that, as was noted in section~\ref{s:TD}, the correlation term $\bkta{\tilde{W}_{\ah};\Thetaa}$ vanishes in an arbitrary process in equilibrium.
To see this first note that the quantity $\bsG{\Theta}+\bGs{\Theta}$ exactly vanishes in an equilibrium process where $\Thetaa(\Ga)=\beta\{H_\nu(\Ga(0))-H_\nu(\Ga(\tau))\}$.
Then the definition \rlb{GG} implies that $G(\Ga)=\odo$, and hence
\eq
\bigl[\Dei(\beta H)(\Ga);G(\Ga)\bigr]_\rho=\ods,
\lb{corvan}
\en
which has no contribution in the quasi-static limit.

\section{Summary and discussions}

In the present paper, we have presented a careful and detailed account of our approach to extend entropy and thermodynamics to nonequilibrium steady states (NESS) in a Markov process describing a heat conducting system.

\paragraph*{Extended Clausius relations and the ``twist'' in thermodynamics of NESS:}
Among various possible routes toward the extended entropy for NESS, we have taken the one which seems to be the closest to the historical introduction of the notion of entropy.
We examined how the Clausius relation $S'_\mathrm{eq}-S_\mathrm{eq}=Q/T=-\sbkt{\Theta}$ (see \rlb{oldCl}), where $\Theta$ is the total entropy production in the heat baths, and the corresponding Gibbs-Shannon expression \rlb{Seq} can be extended to a NESS.
We have then found in \cite{1} that a very natural extension $S'_\mathrm{sym}-S_\mathrm{sym}\simeq-\sbkt{\Theta^\mathrm{ex}}$ (see \rlb{Cl}) was possible, where $\Theta^\mathrm{ex}$ is the excess (or ``renormalized'') version of entropy production in the baths.
The excess entropy production $\Theta^\mathrm{ex}$ characterizes the intrinsic entropy production caused by the operation of the outside agent.

The extended relation is simple and looks natural until this point.
In the present work, however, we found that if we use the same nonequilibrium entropy $S_\mathrm{sym}$ and try to determine this quantity with the precision of $\oes$, we must sometime use the ``nonlinear nonequilibrium'' improvement $S'_\mathrm{sym}-S_\mathrm{sym}\simeq-\sbkt{\Theta^\mathrm{ex}}+\sbkt{\tilde{W};\Theta}/2$ (see \rlb{new}) of the extended Clausius relation, which includes a correlation between the work and the heat.
Although the new relation does not quite look like a thermodynamic relation, we  found that this nonlinear improvement is mandatory and unique once we accept the naturalness of the extended Clausius relation \rlb{Cl}.
Furthermore, it was found that when one wants to determine the difference of entropies in two different NESS with the precision of $\oes$, one may use only the simpler extended Clausius relation without a nonlinear term or must use the nonlinear relation, depending on the operational paths in the parameter space one takes.
The discovery of this annoying but unavoidable ``twist'' in thermodynamics of NESS as well as the new ``nonlinear nonequilibrium'' extended Clausius relation is the main contribution of the present paper.

\paragraph*{Unified derivation of the extended relations:}
From a technical point of view, we have also made a nontrivial progress by  developing a new  method for deriving Clausius-type thermodynamic relations for NESS.
The new method is much more straightforward and transparent than the one we used in the older work \cite{1}.
The key strategy is to  use the simple expression \rlb{exEP} for the excess entropy production along with the representation \rlb{KN} of the probability distribution of NESS.

The new derivation  made it clear that the previously derived extended Clausius relation \rlb{Cl} should be understood as a result within the linear response theory (in a broad sense).
It led us to the improved extended Clausius relation \rlb{new} which explicitly contains the  ``nonlinear nonequilibrium'' term.

It is of course straightforward to extend the present results to other classical systems exhibiting NESS.
In particular models with particle flows can be treated in the line of Section~6 of \cite{JSP}.
Very recently the extended Clausius relation \rlb{Cl} was derived in a quantum system by extending the present method \cite{QM}.
It was found that the nonequilibrium entropy is expressed as the symmetrized von Neumann entropy
\eq
S_\mathrm{sym}[\hat{\rho}_\mathrm{ss}]=
-\frac{1}{2}\,
\mathrm{Tr}\Bigl[\hat{\rho}_\mathrm{ss}\,
\bigl\{\log\hat{\rho}_\mathrm{ss}+
\log(\hat{\mathcal{T}}\hat{\rho}_\mathrm{ss}\hat{\mathcal{T}})
\bigr\}\Bigr],
\en
where $\hat{\mathcal{T}}$ is the density matrix for the NESS and $\hat{\mathcal{T}}$ is the time reversal operator.
We still do not know whether the ``nonlinear nonequilibrium'' relation \rlb{new} can be extended to quantum systems or not.

\paragraph*{Entropy for NESS:}
We have also argued that the symmetrized Gibbs-Shannon entropy \rlb{Sa} is a natural entropy from the point of view of operational thermodynamics.
More precisely if we stick on the natural prescription for ``renormalizing'' the divergent entropy production, this is the unique entropy  to the second order in the ``order of nonequilibrium'' $\epsilon$.

Note that our symmetrized entropy \rlb{Sa} can be seen as the sum of the Shannon entropy of $\rsa(\Ga)$  and the relative entropy between $\rsa(\Ga)$ and its time-reversal $\rsa(\Gas)$.
Therefore the symmetrized entropy certainly reflects the amount of breakdown of the time-reversal symmetry in a NESS.
Unfortunately we still do not have any further interpretation of the new entropy.
It would be quite exciting if one can read off deeper unknown structure of statistical mechanics for NESS from this rather suggestive definition.

A very important issue that must be investigated is to find possible relations between our entropy for NESS and those derived through the studies of large deviation properties \cite{DLS,Rome}.
Although we do not see  any direct connections for the moment, we believe that the thermodynamic-like construction in \cite{Rome} should be investigated in the light of what we have found.

\paragraph*{Future issues:}
There are still many issues in thermodynamics for NESS which must be carefully studied.
The situation about the second law, i.e., the possible inequalities corresponding to the equalities \rlb{Cl} or \rlb{new}, is still rather complicated \cite{2}.

We also note that it is possible to derive a formally exact results which correspond to our thermodynamic relations.
For example, for any process in which the initial and the final states have the same reference inverse temperature \rlb{bref}, we can show that
\eq
F_\mathrm{full}(\alpha')-F_\mathrm{full}(\alpha)=\frac{1}{\beta}
\log\frac{\sbkt{e^{-\Phi_{\ah\dagger}/2}}^{\ah\dagger}}
{\sbkt{e^{-\Phi_{\ah}/2}}^{\ah}},
\lb{Ffull}
\en
where $\ah^\dagger$ is the time-reversal of any path $\ah$, and $\Phi_{\ah}=\Psi_{\ah}+\beta W_{\ah}$.
Here $F_\mathrm{full}(\alpha)$ is the nonequilibrium free energy (to the full order in $\epsilon$) and coincides with the equilibrium free energy in equilibrium states.
The relation \rlb{Ffull}  reduces to (symmetrized version of) the Jarzynski work relation \cite{Chris} when the initial and the final states are in equilibrium.
One can define the nonequilibrium entropy to full order by using the standard formula
\eq
S_\mathrm{full}(\alpha):=\beta\Bigl\{\iG\rsa(\Ga)\,H_\nu(\Ga)-F_\mathrm{full}(\alpha)\Bigr\}.
\en
It is found that this new entropy  coincides with the symmetrized entropy \rlb{Sa} to the precision of $\oes$, but is not exactly the same as \rlb{Sa}.
See \cite{2} for details.
The physical meaning of the suggestive result \rlb{Ffull} is still unclear to us.

So far all of our results are derived using rather formal perturbation in the ``degree of nonequilibrium'' $\epsilon$.
The mathematically minded reader might ask if the results can be formulated as mathematically rigorous statements.
If one reexamines our derivation, it will turn out that most of the procedures are mathematically sound in principle (if one has suitable nice properties about convergence etc.) and the key is whether the representations  \rlb{KN}, \rlb{LR} can be made rigorous.
We believe that these representations can be rigorously stated with some extra work if we make suitable assumptions on the Markov process (see, for example, \cite{LR}).
Such naive rigorous estimates, however, will have very small range of applicability.
It is quite likely that we should take smaller and smaller $\epsilon$ as the system size gets larger.
There is thus no hope of controlling thermodynamic limit with the current technique.
It seems that we need a revolutionary new idea to resolve this issue and obtain meaningful results for large nonequilibrium systems.

Last but not least we wish to stress that both the extended Clausius relation \rlb{Cl} and its ``nonlinear nonequilibrium'' improvements \rlb{new} and  \rlb{newnu} may be investigated experimentally in principle\footnote{
To determine the correlation term in \rlb{newnu}, one needs to measure the heat and the work repeatedly in a series of identically prepared experiments.
}.
It is an exciting challenge to directly observe the ``twist'' in nonequilibrium thermodynamics in experiments.

\bigskip\bigskip


It is a pleasure to thank Herbert Spohn and  Akira Shimizu  for useful discussions in the early stage of the present work, and Giovanni Jona-Lasinio and Keiji Saito for intensive discussions on related issues on NESS.
This work was supported by  grants  Nos.  19540392 (NN), and, 21015005 and 22340109 (SS)  from the Ministry of Education, Science, Sports and Culture of Japan.


\end{document}